\documentclass[%
 reprint,
nofootinbib,
 amsmath,amssymb,
 aps,
prd,
]{revtex4-2}

\usepackage[hidelinks,colorlinks=true,linkcolor=blue,citecolor=blue]{hyperref}
\newcommand\aap{A\&A}                

\newcommand\apjl{ApJ}                
\newcommand\araa{ARA\&A}             
\newcommand\jcap{J.~Cosmology Astropart. Phys.} 

\newcommand\mnras{MNRAS}             

\newcommand{\mathplus}{+}
\usepackage{natbib}
\usepackage{CJK}
\usepackage{graphicx}
\usepackage{dcolumn}
\usepackage{bm}

\usepackage{orcidlink}

\begin{document}


\title{Dissecting environmental effects with eccentric gravitational wave sources}

\author{Lorenz Zwick$^{1,2}$}
\email{lorenz.zwick@nbi.ku.dk}
\author{Kai Hendriks$^{1,2}$}
\author{David O'Neill$^{1}$}
\author{János Takátsy$^{1,2}$}
\author{Philip Kirkeberg$^{1}$}

\author{Christopher Tiede\,\orcidlink{0000-0002-3820-2404}$^{1}$}
\author{Jakob Stegmann$^{3}$}
\author{Johan Samsing$^{1,2}$}

\author{Daniel J. D'Orazio\,\orcidlink{0000-0002-1271-6247}$^{4,5,1}$}
\affiliation{%
 $^{1}$Niels Bohr International Academy, The Niels Bohr Institute, Blegdamsvej 17, DK-2100, Copenhagen, Denmark\\
 $^{2}$Center of Gravity, Niels Bohr Institute, Blegdamsvej 17, 2100 Copenhagen, Denmark.\\
 $^{3}$Max Planck Institute for Astrophysics, Karl-Schwarzschild-Str.~1, 85741 Garching, Germany \\
 $^{4}$Space Telescope Science Institute, 3700 San Martin
Dr., Baltimore, MD 21218, USA \\
$^{5}$Department of Physics and Astronomy, Johns Hopkins University,
3400 North Charles Street, Baltimore, Maryland 21218, USA
}
 %


%


\date{\today}

\begin{abstract}
We model the effect of resonances between time-varying perturbative forces and the epi-cyclical motion of eccentric binaries in the gravitational wave (GW) driven regime. These induce secular drifts in the orbital elements which are reflected in a dephasing of the binary's GW signal, derived here systematically. The resulting dephasing prescriptions showcase a much richer phenomenology with respect to typically adopted power-laws, and are better able to model realistic environmental effects (EE). The most important consequences are for gas embedded binaries, which we analyse in detail with a series of analytical calculations, numerical experiments and a curated set of hydrodynamical simulations for equal masses. Even in these simplified tests, we find the surprising result that dephasing caused by epi-cyclical resonances dominate over expectations based on smoothed or orbit averaged gas drag models in GW signals that retain mild eccentricity in the detector band ($e> 0.05$). We discuss how dissecting GW dephasing in its component Fourier modes can be used to probe the coupling of binaries with their surrounding environment in unprecedented detail.
\end{abstract}

\maketitle

\section{Introduction}
\label{sec:introduction}
A binary source of gravitational waves (GWs) subject to a perturbative force will have the evolution of its orbital elements deviate from the prediction of vacuum General Relativity (GR). For astrophysical sources, this is realised whenever the expected GW driven in-spiral is affected by the binary's residual coupling with its surroundings, resulting in small perturbations to the expected GW emission. These deviations with respect to GR are collectively known as ``environmental effects" (EE) and studying their relevance for both current and future GW detectors has become a rich and growing field \cite{1993chakrabarti,1995ryan,2008barausse,2007levin,kocsis,2014barausse,inayoshi2017,2017meiron,2017Bonetti,2019alejandro,2019randall,2020cardoso,DOrazioGWLens:2020, 2022liu,2022xuan,garg2022,2022cole,2022chandramouli,2022sberna,2023zwick,2023Tiede,2024dyson,2022destounis,2022cardoso,2020caputo,2024zwicknovel,2021andrea,2024basu,2024santoro}. In particular, EEs are emerging as a potentially extremely powerful tool to uncover smoking gun signatures of binary formation channels \citep{2024samsing,kai2024,kai22024,2018PhRvD..98f4012R,liu2015,2021toubiana,2024barandiaran,2024zwick}, and aid population inference based methods in determining the origin of stellar mass black hole (BH) mergers \citep{2002belczynski,2007oleary,2008Sadowski,2016antonini,2017vitale,kavanagh2020,zevin2020,2021zevin,2021kimball,2023santini, 2000ApJ...528L..17P, Lee:2010in,
2010MNRAS.402..371B, 2013MNRAS.435.1358T, 2014MNRAS.440.2714B,
2015PhRvL.115e1101R, 2015ApJ...802L..22R, 2016PhRvD..93h4029R, 2016ApJ...824L...8R,
2016ApJ...824L...8R, 2017MNRAS.464L..36A, 2017MNRAS.469.4665P, Samsing18, 2018MNRAS.tmp.2223S, 2020PhRvD.101l3010S, 2021MNRAS.504..910T, 2013ApJ...773..187N, 2014ApJ...785..116L, 2016ApJ...816...65A, 2016MNRAS.456.4219A, 2017ApJ...836...39S, 2018ApJ...864..134R, 2019ApJ...883...23H, 2021MNRAS.502.2049L, 2022MNRAS.511.1362T,2017ApJ...835..165B,  2017MNRAS.464..946S, 2017arXiv170207818M, 2020ApJ...898...25T, 2022Natur.603..237S,
2023arXiv231213281T, trani2024, Fabj24}. The ways in which EEs can manifest in GW signals is as varied as the physical processes that cause them. A typical example is the source being surrounded by a gaseous medium: Then EEs can arise as a result of frictional forces, migration torques, turbulence or gravitational potentials. These are ultimately sourced by the surrounding mass reservoir, which couples with the binary's energy and angular momentum. Another important case is when the binary is emitting GWs in the vicinity of a third massive body. Then, tidal forces, lensing and center of mass accelerations will similarly produce a plethora of EE that induce changes in the radiated GWs.

Despite all of the complexity, or perhaps because of it, the most studied tracer of EEs in terms of GW observables is a relatively simple quantity: the so-called dephasing of a signal. Dephasing refers to a small deviation $\delta \phi$ from the expected binary phase evolution, that accumulates over the entire observation. In almost all works, dephasing is modelled in terms of two independent parameters, an amplitude $A$ and a power law index describing the accumulation rate:
\begin{align}
    \delta \phi \sim A \times f^{\alpha}
\end{align}
where $f$ is the GW frequency, and both $A$ and $\alpha$ are related to the EE in question. Reconstructing the parameters $A$ and $\alpha$ in an observation would, in principle, allow determination of characteristics of the source's environment, such the average density of a surrounding medium or the mass/distance of a nearby third body. However, the detection of dephasing (and EE in general) in realistic signals is considered an optimistic prospect, as the more skeptically minded maintain that such measurements would only be possible for future GW detectors, and even so only for a small subset of ``golden" sources. The reasons for this are varied and legitimate, and relate to both the astrophysics of GW sources and the constraints of GW data analysis \citep[see e.g.,][]{2023Keijriwal,owen23}. Nevertheless, the scientific potential of detecting EEs along with the indications that GW sources form in complex high density environments, strongly justify the rising interest in the topic. In fact, the most recent estimates indicate that EE should be detectable in hundreds of stellar mass binary sources of third generation GW detectors \citep{2025zwick}.

Are there any neglected aspects of EEs that could be even more promising in terms of prospective detections? Focusing more specifically on dephasing: Are there any additional characteristics of the typical environment of GW sources that have been previously neglected or modelled too simplistically? And given a detection of dephasing: Are there ways in which more information about the source's environment can be extracted, other than the one provided by the rather limited parameters $A$ and $\alpha$? In a recent article \citep[][hereafter PI]{2024zwick} we highlighted how in the majority of astrophysical scenarios EEs are in fact much more complex than what is typically modelled. In particular, both binaries embedded in accretion discs or perturbed by third bodies are subject to highly variable forces composed of numerous Fourier modes. These oscillating force components are often neglected when modelling perturbative binary dynamics, under the justification that they typically do not produce any long term drift in the binary's evolution\footnote{See Ref.~\cite{2025copparoni} for an analysis of the consequences of using simplified prescriptions when attempting to recover dephasing from stochastic hydrodynamical torques.}. However, as shown in both hydrodynamical simulations of gas embedded binaries and direct integrations of three body systems, it is the case that the typical variability of the perturbative forces is significantly larger than its smoothed or orbit averaged magnitude, sometimes by factors of $\sim 10$ to $100$ \citep{KleyNelson:Review:2012,Roedig_Trqs+2012,2021andrea,tiede2020,2024zwick}.


In PI, (and previously in \cite{2022zwick}) we presented a first analysis of such time variable perturbations, focused on the resulting direct observables in the GW phase of inspiralling binaries. Here, we instead focus on the effects of resonances between the time-varying perturbative forces and the epi-cyclical motion of eccentric inspiralling binaries (which was noted but not studied in PI). We present a systematic way to dissect the various resonant contributions to the total secular evolution of the orbital elements, which in turn are reflected in a dephasing of the binary's GW signal. Crucially, certain components of the dephasing are caused by modes of the perturbative force that vanish when an orbit average is performed. We find the surprising result that, for sufficiently eccentric sources, dephasing caused by such resonances will typically be larger than expectations based on smoothed or orbit averaged EEs, and we lay out the consequences of this fact in detail. These results are timely, considering  the growing number of detections of binary mergers that show evidence for non-zero eccentricity \citep{2021ApJ...921L..31R,2024gupte}. As detectors improve, it will be crucial to take the results here presented into account for the analysis of EE in eccentric signals for both future space-borne milli-Herz detectors \citep{2022xuan,2023PhRvD.107d3009X}, upcoming Ligo/Virgo/Kagra upgrades and third generation ground based detectors \cite{2020maggioreet,2023arXiv230613745E}.

The paper is structured as follows: In section \ref{sec:Resonances}, we derive and solve the evolution equations of an eccentric binary perturbed by time-varying forces, while the resulting dephasing is derived in section \ref{sec:Dephasing}. In section \ref{sec:phenom}, we discuss the phenomenology of the resulting dephasing prescriptions. Sections \ref{sec:astro} and  \ref{sec:astro_numerical} are devoted to demonstrating the existence and relevance of time-varying forces and the resulting dephasing in realistic astrophysical systems. In the former, we analyse some commonly adopted analytical EE models, while in the latter we analyse gas-embedded binaries in detail with a series of semi-analytical and numerical experiments. We provide a short summary of the results in section \ref{sec:conclusion}.

\begin{figure}
    \centering
    \includegraphics[width=0.85\linewidth]{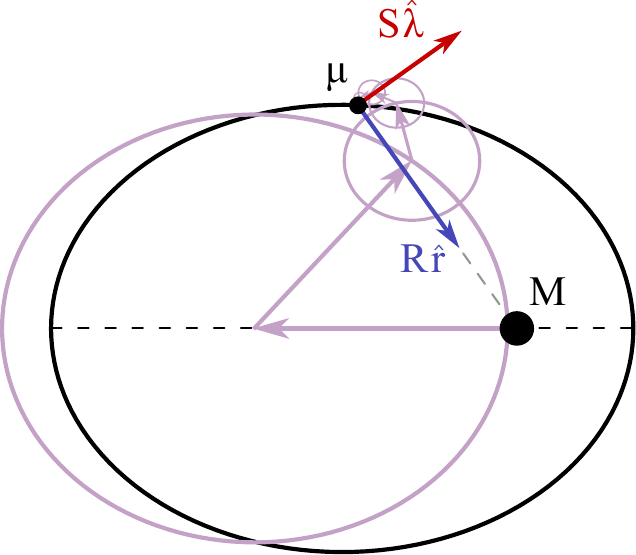}\\
    \vspace{0.4cm}
    \includegraphics[width=1\linewidth]{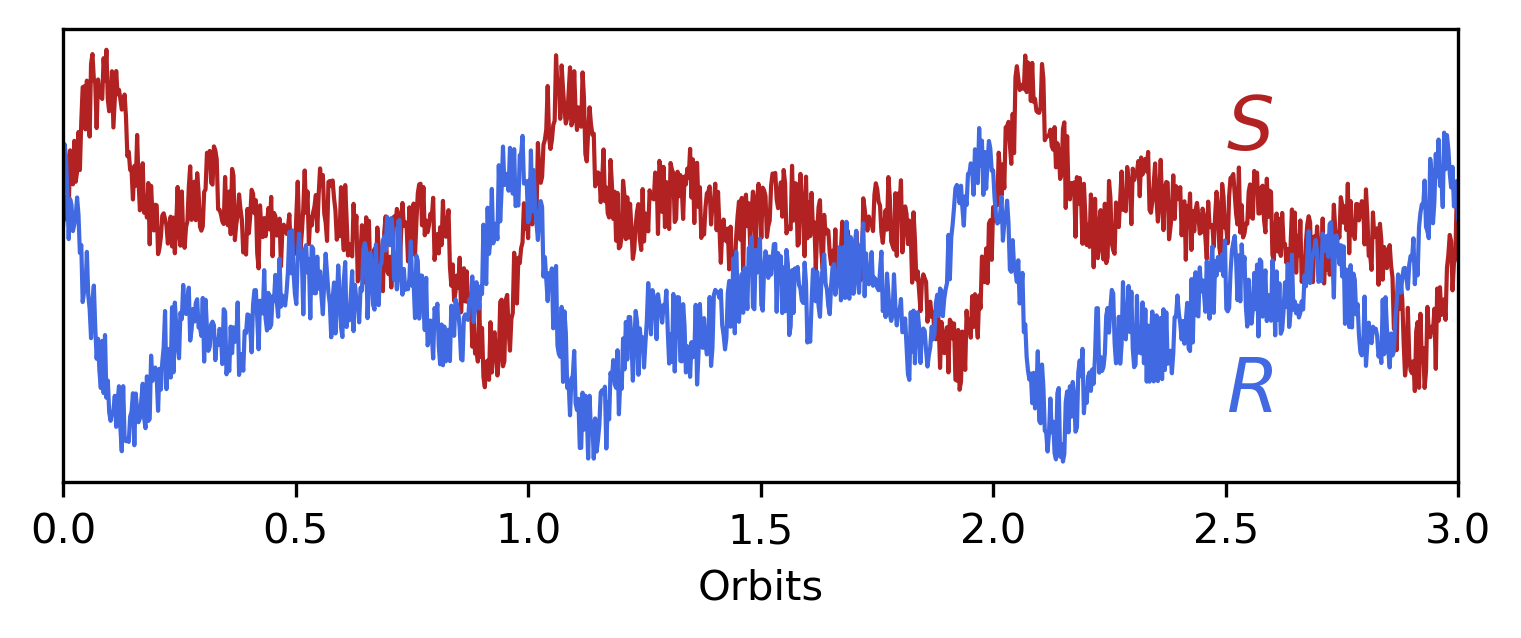}
    \caption{Illustration of the principal ingredients used to derive the results of this work. A binary on an eccentric orbit (black orbit) is perturbed by a azimuthal (red) and radial (blue) force with a complex Fourier spectrum at multiples of the orbital frequency. Through the Lagrange planetary equations, we derive how the $n$-th force Fourier components enter in resonance with the $n$-th epicyclical correction to the orbital motion, here all represented by the pink ellipses.}
    \label{fig:Illustration}
\end{figure}


\section{Perturbative Forces and Epicyclic Resonances}
\label{sec:Resonances}
\subsection{Forces and Orbital Elements}
Our goal is to build up a systematic way to describe perturbations to the evolution of the orbital elements of an eccentric binary, such that we may clearly distinguish the contributions to the GW dephasing. A basic sketch of the ingredients needed to derive the results of this work is shown in Fig. \ref{fig:Illustration}, and the methodology we adopt is also described in PI \citep[see also e.g.][]{2014poisson,2014will,2017will,2018ApJ...863...68L}. We wish to describe time-varying perturbative force (here expressed in units of acceleration) ${\rm{d}} \mathbf{F}$ on a binary system. The perturbative force is decomposed into radial, azimuthal and normal components (following \cite{2014poisson}):
\begin{align}
    {\rm{d}} \mathbf{F} = R \mathbf{\hat{r}} + S \mathbf{\hat{\boldsymbol\lambda}} + N \mathbf{\hat{z}},
\end{align}
We are only interested in modeling in-plane forces acting on a Newtonian binary as these correspond to the first order perturbations that can directly affect the phase of an inspirallng GW source. {While the Newtonian assumption is known to break down for relativistic binaries, the perturbative treatment is still justified because any additional cross terms between EEs and relativistic dynamics are suppressed with respect to the leading order terms in $1/c^2$ \citep[see e.g.][]{2017will,2024zwick}. In terms of parameter estimation, the mixing of EEs and PN effects typically only becomes relevant at 5th or 6th PN order \citep{2023zwick}, which means that here we expect any significant driver of PN phase evolution to add linearly to our results (with the possible exception of interactions between precession and the force components discussed in section \ref{sec:phenom}.)} We express the in-plane forces as a Fourier series in terms of the binary's Keplerian frequency $f_{\rm K}$ and the time $t$:
\begin{align}
    \label{eq:perturbingforces}
   {\rm{d}} \mathbf{F}^{\rm p} &= \text{Re}\Bigg(\sum_n \Big[ B_{n}^{\rm S} \exp\left(i2 \pi   n f_{\rm K} t\right) \mathbf{\hat{\boldsymbol\lambda}} \nonumber \\ &+ B^{\rm{R}}_n \exp\left(i2 \pi  n f_{\rm K} t\right) \mathbf{\hat{r}}\Big]\Bigg),
\end{align}
Where the complex coefficients $B^{\rm T}_n$ and $B^{\rm R}_n$ are in general functions of the orbital elements and both the amplitudes and the phases must be supplied. The goal of this representation is to clearly distinguish between constant and time-variable force components, which will enable us to separate the various contributions to the total dephasing. It is based on two crucial assumptions: The first one is the adiabatic approximation, i.e. the limit in which the orbital elements evolve slowly with respect to the orbital timescale. The second one is that super-orbital variability is periodic in the binary orbit, i.e. that fast force variability is related to binary motion itself. In other words, we operate in a two-timescale framework in which slow super-orbital perturbations are described in the adiabatic evolution of the coefficients, while fast sub-orbital perturbations are encoded in the instantaneous value of high $n$ coefficients. Therefore, for most applications in this work we will have:
\begin{align}
    B^{\rm{R}}_n &= B^{\rm{R}}_n(a,e) \\
    B_{n}^{\rm S} &= B_{n}^{\rm S}(a,e),
\end{align}
where the parameter $n$ in Eq. \ref{eq:perturbingforces} denotes the given multiple of the orbital frequency. It is crucial to note, that this form parametrisation effectively implies that the adiabatic approximation holds throughout. Beyond these assumptions however, Eqs. \ref{eq:perturbingforces} are very general and can describe a a wide range of external perturbations, which are described in section \ref{sec:phenom}.

In this work we are interested in following the binary's semi-major axis $a$ and eccentricity $e$, as these control the secular evolution of the binary inspiral. Note that pericenter precession can also influence the binary evolution and result in additional dephasing \citep{2023Tiede,2022chia,1994karas}, though we do not model this here. The Lagrange planetary equations for the latter two elements read \citep[e.g.,][]{Merritt2013}:
\begin{align}
\label{eq:Lagrange_p}
    \frac{{\rm d}p}{{\rm d}t} &= 2 \sqrt{\frac{p^3}{G M}} \frac{S}{\left(1 + e \cos \nu \right)} \\
    \label{eq:Lagrange_e}
    \frac{{\rm d}e}{{\rm d}t} &= \sqrt{\frac{p}{GM}}\left[R \sin \nu + \frac{2 \cos \nu + e(1+\cos^2 \nu)}{1+e \cos \nu}S \right]
\end{align}
where $p = a(1-e^2)$ is the semi-latus rectum and $\nu$ is the true anomaly. The binary in space is described by the following vector:
\begin{align}
    r(\nu)\left\{ \cos\left( \nu + \omega \right),\sin\left( \nu + \omega \right) \right\},
\end{align}
where $\omega$ is the argument of pericenter and the radial separation is:
\begin{align}
    r(\nu) = \frac{a(1-e^2)}{1 +e\cos(\nu)}.
\end{align}

\subsection{Perturbations of Eccentric Binaries}
As performed in PI, it is possible to solve Eqs. \ref{eq:Lagrange_p} and \ref{eq:Lagrange_e} analytically in first order perturbation theory, for a circular orbit. Here we extend the calculations to mildly eccentric orbits by means of an analytic expansion of the equation of the center. We start with the definition of the mean anomaly ${\rm{An}}$:
\begin{align}
    {\rm An} = \sqrt{\frac{G M}{a^3}}t,
\end{align}
and:
\begin{align}
    \nu &= {\rm{An}} + 2 \sum_{s=1}^{\infty}\frac{1}{s}\bigg[ J_{s}(se) \nonumber \\&+\sum_{p=1}^{\infty}\frac{1}{e}(1-\sqrt{1-e^2})\left[J_{s-p}(se) + J_{s+p}(se)\right] \bigg]
\end{align}
Then, the Kepler problem can be expanded around small eccentricities, yielding the following approximation for the true anomaly as a function of time:
\begin{align}
\label{eq:centers}
    \nu(t) &\approx {\rm An} + 2e \sin\left({\rm An} \right) + \frac{5e^2}{4}\sin\left(2{\rm An} \right) \nonumber \\ &+ \frac{e^3}{12}\left[13\sin\left(3{\rm An} \right) - 3\sin\left({\rm An}\right) \right] \nonumber \\ &+ \frac{e^4}{96}\left[103\sin\left(4{\rm An} \right) - 44\sin\left(2{\rm An} \right) \right] \nonumber \\
    &+ \mathcal{O}\left[e^5\right],
\end{align}
where we see the various epi-cyclical modifications to circular orbital motion. Inserting Eq. \ref{eq:centers} in the Lagarange planetary equations and integrating directly with symbolic manipulation software \texttt{Mathematica} yields analytical solutions for the perturbed orbital elements $a_{\rm p} = a_0 + \delta a$ and $e_{\rm p} = e_0 + \delta e$, where the subscript ``0" denotes the unperturbed elements that evolve adiabatically in vacuum. As detailed in PI, 
the solutions to $\delta a$ and $\delta e$ are given in terms of Fourier series, of similar form to Eqs. \ref{eq:Lagrange_p} and \ref{eq:Lagrange_e}, that are too long to report explicitly. In addition to oscillating components, particular combinations of phases and values of $n$ give rise to orbital resonances, which induce secular drifts. The latter can be identified by performing an orbit average:
\begin{align}
    \dot{a} \rvert _{\rm avrg} &= f_{\rm K}\int_0^{1/f_{\rm K}} \dot{a}(t) \, {\rm{d}}t \\
    \dot{e} \rvert _{\rm avrg} &= f_{\rm K}\int_0^{1/f_{\rm K}}  \dot{e}(t) \, {\rm{d}}t.
\end{align}
This procedure suppresses the vast majority perturbations arising from non vanishing coefficients $B_{n}^{\rm R}$ and $B_{n}^{\rm S}$. Indeed, only the force Fourier modes that are in resonance with the various epi-cyclical contributions to the equation of centers (Eq. \ref{eq:centers}) survive the averaging. Hereafter, we focus solely on such secular drifts and do not explicitly specify the averaged quantities.  We note here that the impact of orbital resonances in GWs has been studied in \cite{2021speri,2024areti} in the context of vacuum inspirals for extreme mass ratio (EMRI) sources, and in the specific case of tidal forces in \cite{2019Bonga,2024camilloni,cocco} as a probe of the strong gravity regime. It has also been studied as a potential way to probe GW backgrounds in \cite{2022blas}.

\subsection{Secular orbital evolution for mildly eccentric binaries}
Here we report the explicit solutions to the secular evolution of the orbital elements $a$ and $e$ for a binary perturbed by a force of the form shown in Eq. \ref{eq:perturbingforces}. To order $e_0 ^2$ they read:
\begin{align}
\label{eq:adot}
    \dot{a}_{\rm p} \sqrt{\frac{GM}{p_0^3}} &= 2 B_{0}^{\rm S} + e_0\left(B_{1}^{\rm S} + iB_1^{\rm R} \right) \nonumber \\ & + e_0^2\left(  B_{2}^{\rm S}+ 2 B_{1}^{\rm S} +  iB_2^{\rm R}\right) + \mathcal{O}\left( e_0^3\right) \\
    \dot{e}_{\rm p}\sqrt{\frac{GM}{p_0}} &= B_{1}^{\rm S} + i\frac{1}{2}B_1^{\rm R} \nonumber \\ &+ \frac{1}{4}e_0\left( - 6B_{1}^{\rm S}+ 3B_{2}^{\rm S} + 2 i B_2^{\rm R}\right) \nonumber \\
    &+ \frac{1}{16}e_0^2(-7 i B^{\rm R}_1 + 9 i B^{\rm R}_3 - 12 B_{1}^{\rm S}  \nonumber \\&+ 12 B_{3}^{\rm S})+ \mathcal{O}\left( e_0^3\right),\label{eq:edot}
\end{align}
where only the real part contributes. The solutions up to $e_0^4$ are reported in the appendix, and used for numerical calculations in later sections. Secular drift of the semi-major axis is produced trivially by the constant force component $B_{1}^{\rm S}$. Epi-cyclical resonances pick out additional contributions from the $n=1$, $n=2$ and even $n=3$ modes, provided that eccentricity is present and that the perturbations have the appropriate complex phase. Crucially, we note again how these particular contributions to the overall secular drift of $a$ and $e$ arise as a consequence of force components that vanish when orbit averaged. We also note that expanding to higher orders in the eccentricity picks out additional resonances from higher modes, as seen in the appendix. Computations up to arbitrary order are possible, though not useful here, and would contain resonances with arbitrarily high values of $n$. Finally, note also how radial and azimuthal perturbations add up linearly when out of phase by 90$^{\circ}$.
\newline \newline 

Before continuing, we briefly highlight the conceptual insight that led us to present the calculations above. Consider as an example the operation of performing the orbit average of the energy flux $<\dot{E}>$, where the brackets denote the orbit average and $\dot{E}= {\rm d}{\mathbf{F}} \cdot \mathbf{v}=T\times v_{T}$, where $T$ is the tangential force and $v_{\rm T}$ the tangential velocity. In this paper, we are exploring in detail the consequence of the fact that:
\begin{align}
    <T\times v_{T}> \neq <T> \times <v_{T}> .
\end{align}
While this statement is perhaps obvious, the ways in which the equality fails is non-trivial for binaries that experience forces  dominated by variability. We are presenting a systematic way to characterise and quantify the consequences of such failures in terms of GW observables. In the following, we will show how these insights can lead to a more sophisticated modelling of the phase evolution of perturbed sources of GW, and potentially new predictions for the de-phasing in the expected GW signal of astrophysical binary sources.

\section{Dephasing of GW signals}
\label{sec:Dephasing}
\subsection{Basics of the GW phase}
Here we go through some basic elements required to compute the phase evolution of GW sources at Newtonian order, before highlighting the new elements introduced in this work. More thorough derivations can be found in e.g. Refs. \cite{1994cutler,2018maggiore,kocsis,2014barausse,2023zwick,garg2022}. The frequency of a binary's orbit $f_{\rm K}$ is related exclusively to its semi-major axis $a$, leading to the following relation valid at Newtonian order:
\begin{align}
\label{eq:adotGW}
  \frac{\dot{a}}{a} = -\frac{3}{2}\frac{\dot{f}_{\rm K}}{f_{\rm K}}.
\end{align}
Additionally, the GW frequency $f$ is related to the binary orbital frequency as $f = 2 f_{\rm K}$ at quadrupolar order. In vacuum, binaries decay due to the emission of GWs:
\begin{align}
    \dot{a}_{\rm vac} &= \dot{a}_{\rm vac}^{e=0} \times F(e), \\
    F(e) &= \left(1+\frac{73}{24}e^2 + \frac{37}{96}e^4 \right)(1-e^2)^{-7/2},
\end{align}
where we use the classic result of Peters \& Mathews \citep{peters1964}. The total accumulated GW phase of a binary is then found by integrating the frequency over the binary's chirp:
\begin{align}
    \phi_{\rm GW} = 2 \pi \int \frac{f}{\dot{f}}\, {\rm{d}}f,
\end{align}
where:
\begin{align}
    \label{eq:chirpf}
    \dot{f}_{\rm vac} = \frac{96}{5} \pi ^{8/3}\left(\frac{G \mathcal{M}_z}{c^3}\right)^{5/3}f^{11/3} \times F(e),
\end{align}
and $\mathcal{M}_z$ is the binary's redshifted chirp mass. For eccentric binaries, the frequency evolution is coupled to the eccentricity by means of the following relation:
\begin{align}
\label{eq:ae}
    a(e) = a_{\rm in}\frac{g(e)}{g(e_{\rm in})},
\end{align}
where \citep{peters1964}:
\begin{align}
\label{eq:ge}
    g(e) = \frac{e^{12/19}}{1 - e^2}\bigg(1 + \frac{121}{304}e^2\bigg)^{870/2299}.
\end{align}
For later convenience, we note that Eq. \ref{eq:ae} can be inverted at arbitrary orders in eccentricity \citep[see e.g.][]{2009yunes}:
\begin{align}
    \label{eq:eevol}
    e(a) &\approx \left( g(e_{\rm in})\frac{a}{a_{\rm in}}\right)^{19/12} - \frac{3323}{1824}\left(g(e_{\rm in})\frac{a}{a_{\rm in}} \right)^{19/4} \nonumber \\
    &+ \mathcal{O}\left(\left(g(e_{\rm in})\frac{a}{a_{\rm in}}\right)^{95/12} \right),
\end{align}
where the subscript ``in" denotes the initial values at some specified time and we recall that $a\propto f^{-3/2}$.
The above equation is accurate to high order $\mathcal{O}(e^{88/19})$, and will be the basis of our analytical results. However, it presents some unwanted features such as a turn-over for high values of the eccentricity. For applications to highly eccentric binaries, one can instead invert the equation numerically. This will also be the procedure we adopt in our numerical studies.
\subsection{Calculation of the GW dephasing}
The perturbations in the orbital elements computed in Eqs. \ref{eq:adot} and \ref{eq:edot} are reflected in the GW phase in two separate ways. Firstly, a perturbation in the semi-major axis evolution $\dot{a}_{\rm p}$ directly causes a corresponding perturbation to the binary's frequency evolution in vacuum, yielding:
\begin{align}
    \dot{f}_{\rm GW} \to  \dot{f}_{\rm vac} - \frac{2}{3}\frac{\dot{a}_{\rm p}}{a}f.
\end{align}
Secondly, a perturbation in the eccentricity evolution indirectly causes a corresponding perturbation to the binary's frequency evolution through Eq. \ref{eq:dephasing_master}, yielding:
\begin{align}
    \dot{f}_{\rm GW} \to  \dot{f}_{\rm vac}  + \frac{{\rm d} \dot{f}_{\rm{vac}} }{{\rm d} e_0}(e_0)\delta e,
\end{align}
where we expanded in small perturbations, and:
\begin{align}  
    \frac{{\rm d} \dot{f} }{{\rm d} e_0} = \dot{f}_{\rm GW}^{e_0=0}e_0\frac{1256 + 1608e_0^2 + 111e_0^4}{96(1 - e_0^2)^{9/2}}.
\end{align}
Thus, the total phase of the perturbed binary can be computed as:
\begin{align}
    \phi_{\rm tot} &= 2 \pi \int f \nonumber \\ &\times\left(\dot{f}_{\rm{vac}} + \dot{f}_{\rm{p}} + \frac{{\rm d} \dot{f}_{\rm{vac}} }{{\rm d} e_0}\delta e(f)\right)^{-1} \, {\rm{d}}f,
\end{align}
where here $\delta e$ is the change in eccentricity accumulated from the start of the observation. Defining the dephasing of the GW as:
\begin{align}
     \phi_{\rm tot}\approx \phi_{\rm vac} + \delta \phi,
\end{align}
and expanding to first order in small perturbations we find:
\begin{align}
\label{eq:dephasing_master}
    \delta \phi = - 2\pi \int \frac{f}{\dot{f}_{\rm{vac}}^2} \left(\dot{f}_{\rm p} + \frac{{\rm d} \dot{f} }{{\rm d} e_0} \delta e\right) \, {\rm{d}}f,
\end{align}
where $\dot{f}_{\rm p}$ is the chirp induced directly from the semi-major axis perturbations. The precise meaning of $\delta e$ is the difference of eccentricity between the perturbed and unperturbed binary at a given frequency $f$, which we set here to be zero at merger. This quantity is thoroughly discussed in \cite{pedo} (though see also \cite{2020deme,2022chandramouli}) and is determined by the interaction of both eccentricity and semi-major axis perturbations, as well as the relation between the two in the GW driven regime (Eq. \ref{eq:ge}). It reads:
\begin{align}
    \delta e = \frac{g(e_0)}{g'(e_0)}\int \frac{{\rm d}f}{\dot{f}_{\rm vac}} \left[ \frac{g'(e_0)}{g(e_0)} \dot{e}_{\rm p} - \frac{\dot{a}_{\rm p}}{a_0}\right],
\end{align}
which at lowest order in eccentricity reduces to:
\begin{align}
    \delta e = e_0(f)\int \frac{\dot{e}_{\rm p}(f')}{\dot{f}_{\rm vac}^{{e=0}}(f') e_0(f')} \, {\rm d}f',
\end{align}
{where $e_0(f)$ denotes the unperturbed eccentricity, which decays as a function of $f$ due to the emission of GW. Note that in this paper we will express results in terms of both $e_0(f)$, i.e. the eccentricity of the binary over the binary chirp, as well as $e_0(f_0)$, i.e. the eccentricity of the binary at a given reference frequency $f_0$.} The perturbation $\delta e$ is already an integrated quantity and in the general case the total dephasing can pick up contributions that, schematically, are of the form $\int {\rm d}f$ as well as $\int(\int {\rm d}f){\rm d}f$. This can interact with the particular form of the Fourier coefficients in interesting ways, as explored in Section \ref{sec:phenom}.
In combination with Eqs. \ref{eq:adot} and \ref{eq:edot}, Eq. \ref{eq:dephasing_master} provides us with a tool to relate a set of Fourier coefficients $B_n^{\rm R}$ and $B_n^{\rm R}$ of any perturbative force acting on the binary (see Eq. \ref{eq:perturbingforces}) with a resulting dephasing in the GW wave. In particular, one can build up the total dephasing in the following way:
\begin{align}
    \delta \phi = \rm{Re}\left(\sum_n \delta \phi_n,\right),
\end{align}
where each component $\delta \phi_n$ exclusively depends on the coefficients with the same value of $n$, and once again we are careful to only select the real part. We will use this equation to numerically compute the dephasing caused by oscillating forces for some exemplary astrophysical cases in section \ref{sec:astro}. Additionally, it shows how a more thorough treatment of peturbative forces and their effect on inspiralling binaries can lead to a rich phenomenology of dephasing prescriptions. Before concluding this section, we briefly mention that the binary phase $\phi$ and the corresponding phase of the GW waveform $\Psi$ differ conceptually \cite{1994cutler,pedo}. The dephasing in the Fourier waveform is given by the following relation:
\begin{align}
    \frac{{\rm d}^2\delta \Psi}{{\rm d}f^2} \approx -2 \pi\frac{\dot{f}_{\rm p }}{\dot{f}_{\rm vac}},
\end{align}
which provides a way to relate the results in this work to GW observables more directly. We also note here that all of the calculations above are performed in the limit of a Newtonian binary. While this is correct as a first order approximation, it has to be verified explicitly that the resulting dephasing prescriptions can be adopted one-to-one for relativistic precessing orbits in the stationary phase approximation (see e.g. \cite{2022chandramouli}). We leave a more thorough analysis of this for future work.


\section{Phenomenology and Order of Magnitude Estimates}
\label{sec:phenom}
\begin{figure*}
    \centering

\includegraphics[width=0.66\columnwidth]{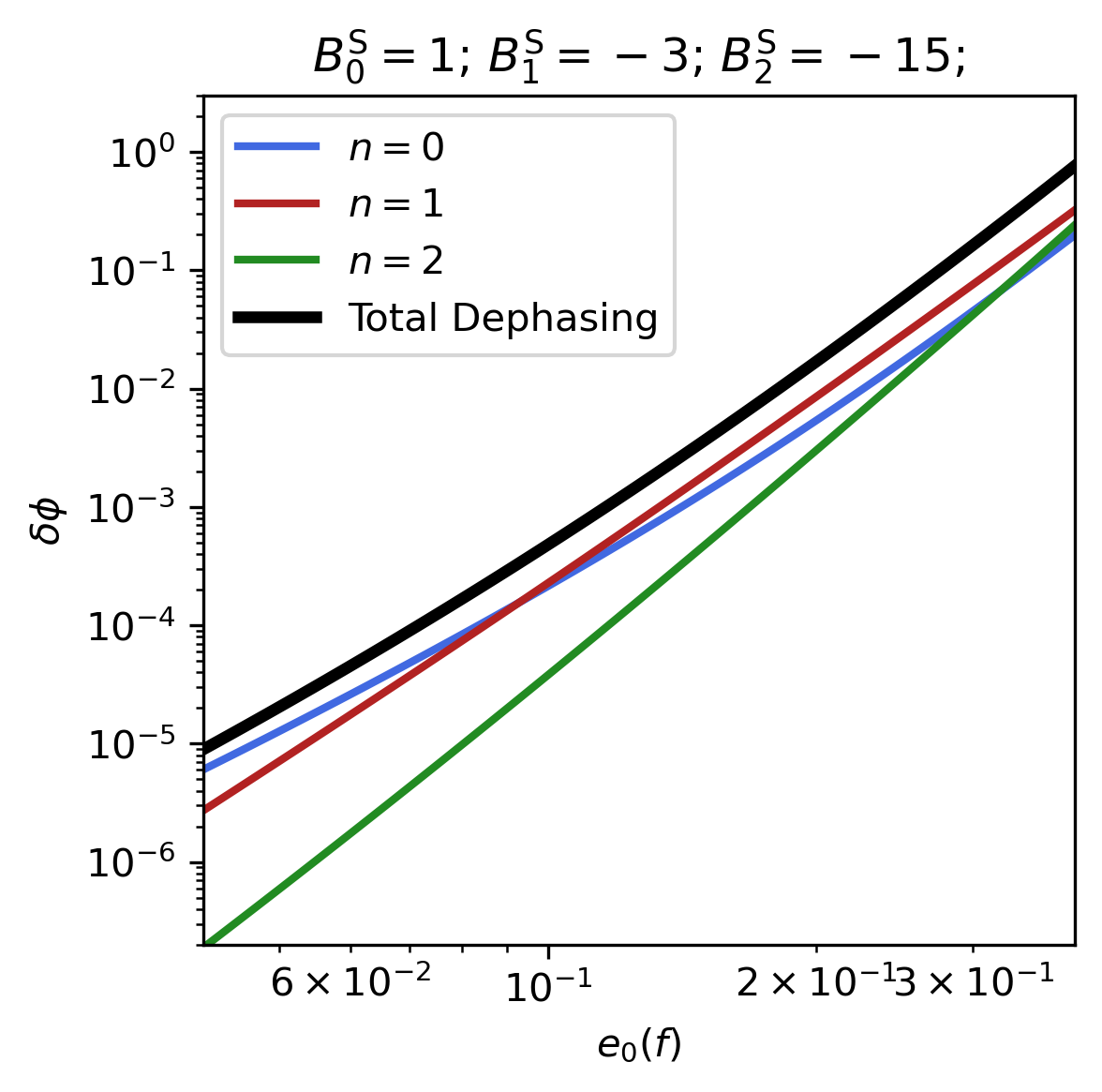}
\includegraphics[width=0.66\columnwidth]{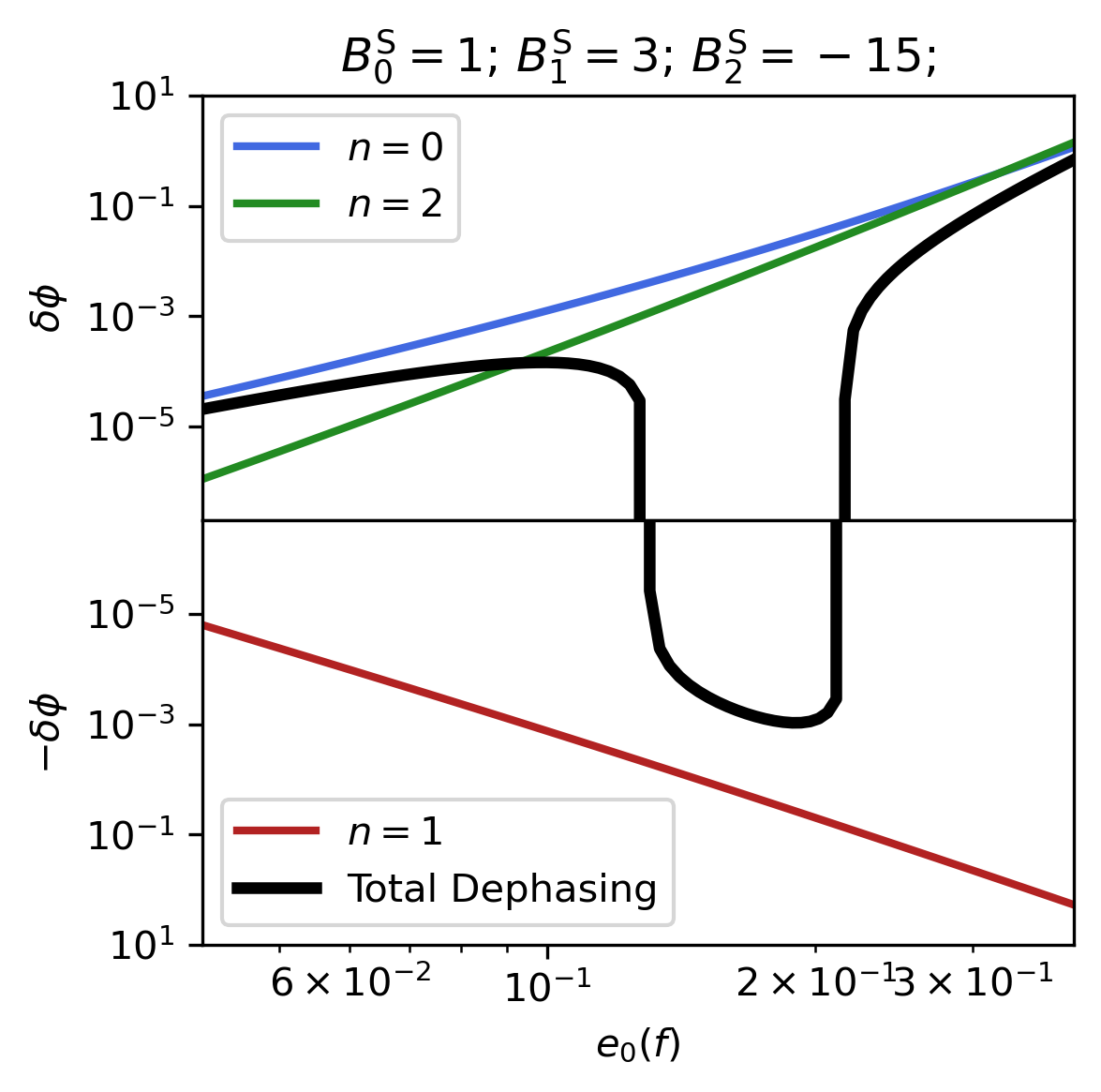}
\includegraphics[width=0.66\columnwidth]{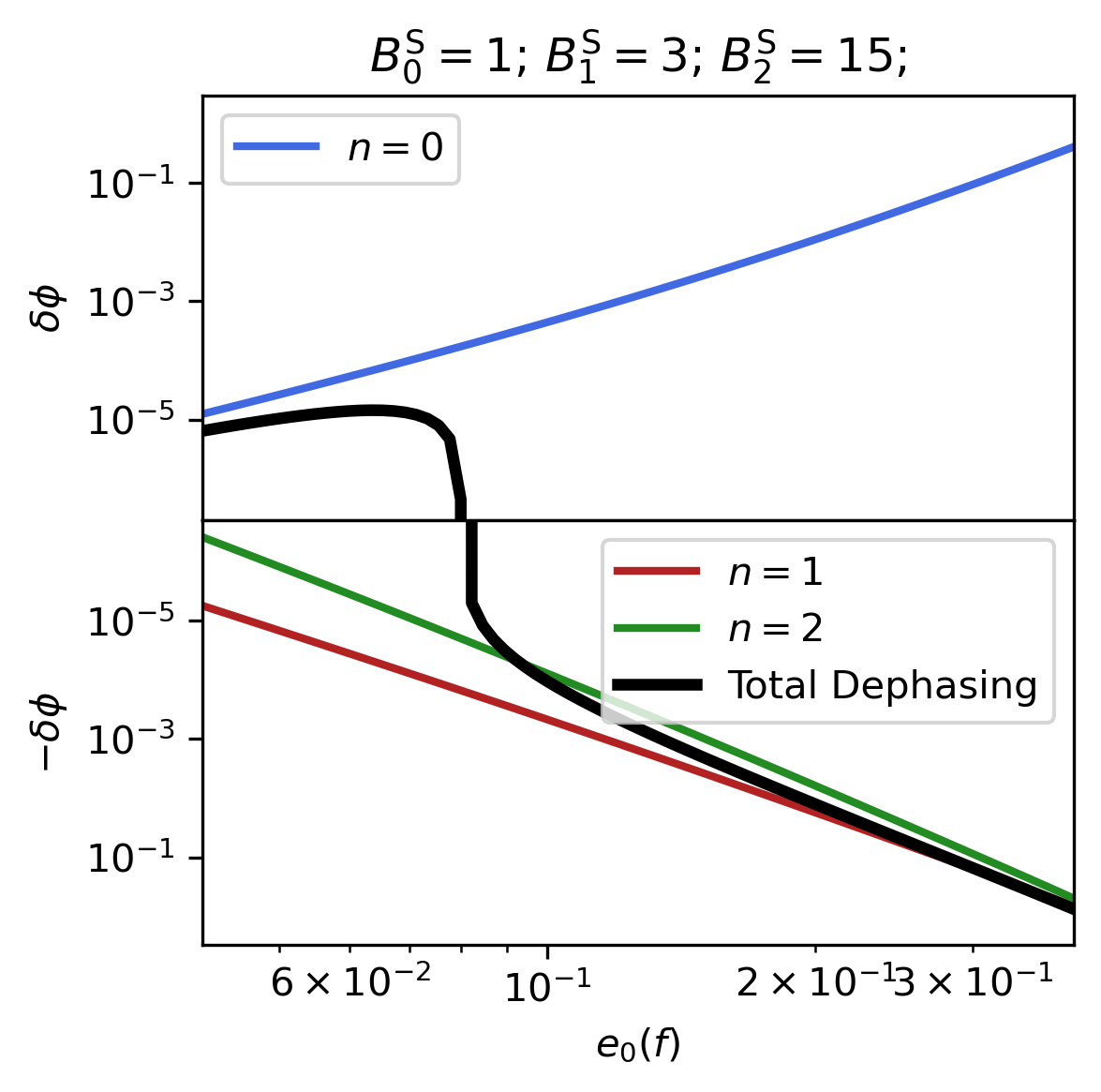}

\caption{\label{fig:phenomenology_constant}Various possible dephasing curves in the GW of a perturbed, chirping eccentric binary, as a function of eccentricity. The binary is perturbed by a normalised force with $\lvert B_{1}^{\rm S} \rvert=1$, $\lvert B_{1}^{\rm S} \rvert=3$ and $\lvert B_{2}^{\rm S} \rvert=15$, here with arbitrary units and where the radial forces have been set to zero. The phases of the Fourier components determines whether the induced dephasing from different epicyclical resonances (coloured lines) adds coherently. The result of changing the phases is to produce a wide range of possible behaviours, shown in the three panels. The plots can be rescaled by varying masses and reference eccentricity. An eccentric source entering a detector band with non negligible eccentricity can showcase a complex dephasing phenomenology.}
\end{figure*}

\begin{figure*}

    \centering

\includegraphics[width=0.99\columnwidth]{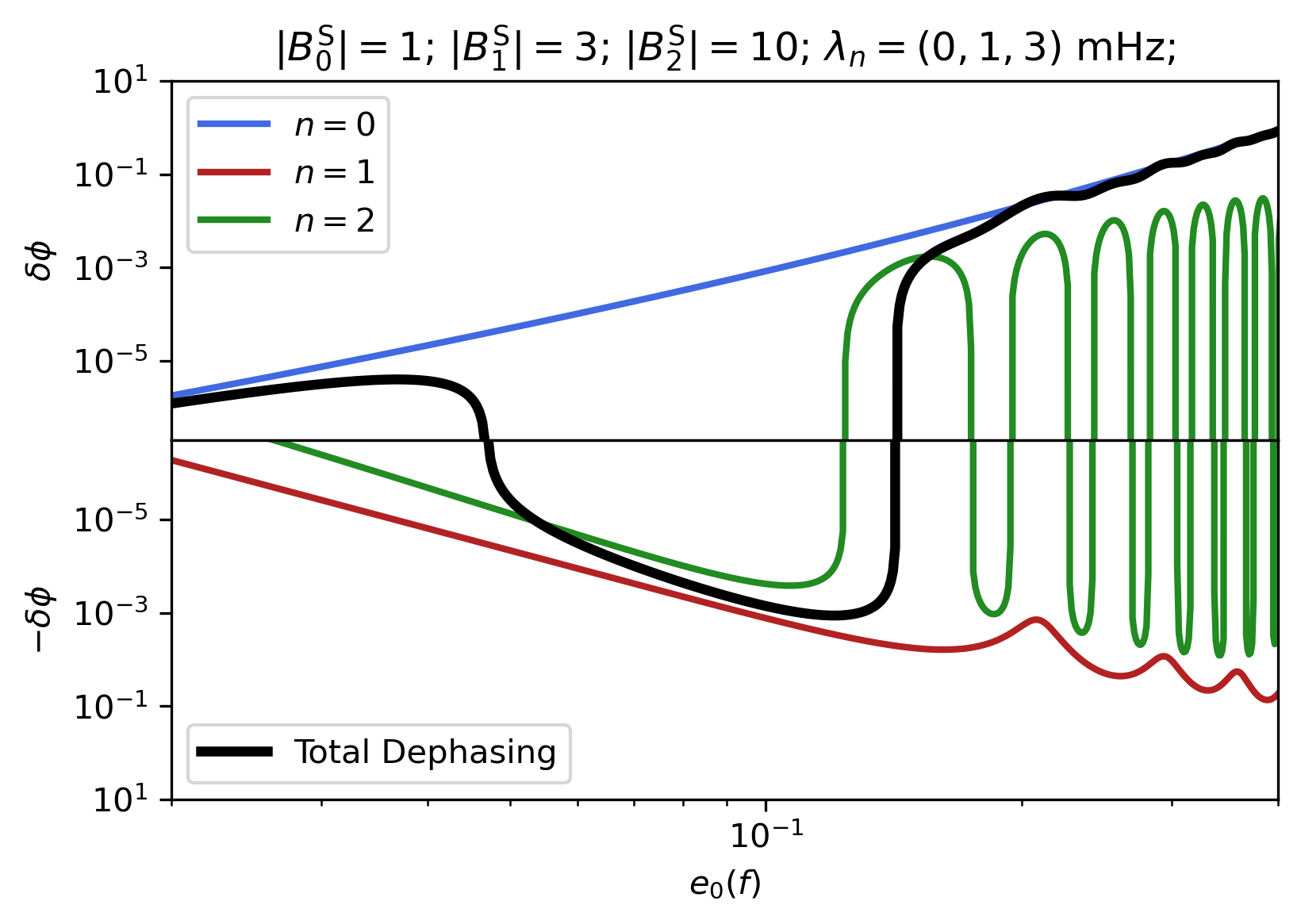}
\includegraphics[width=0.99\columnwidth]{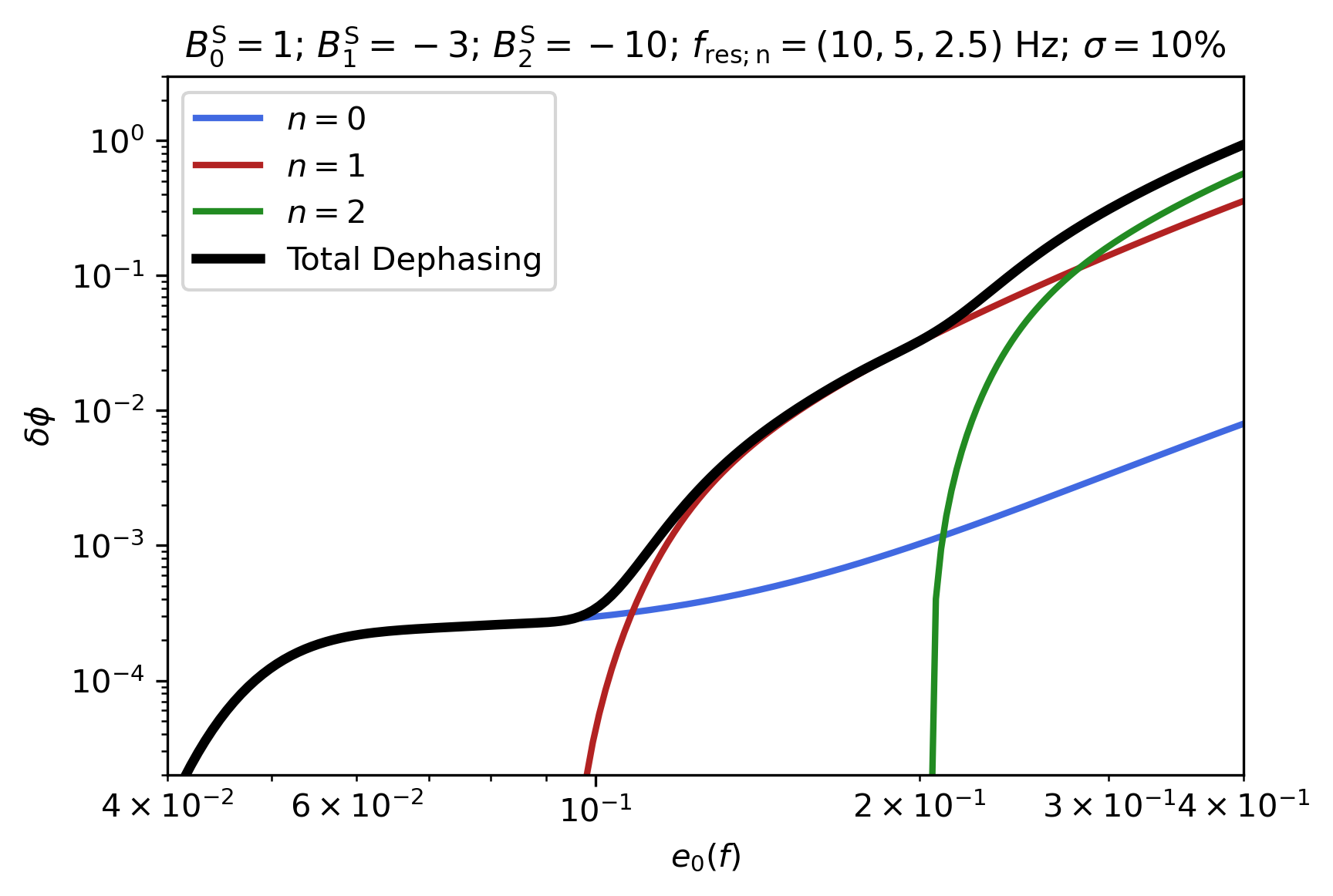}

\caption{\label{fig:phenomenology_other}Dephasing in the GW of a perturbed, chirping eccentric binary system, as function of the eccentricity. In the left panel, the binary is perturbed by a normalised force with $\lvert B_{0}^{\rm S} \rvert=1$, $\lvert B_{1}^{\rm S} \rvert=3$ and $\lvert B_{2}^{\rm S} \rvert=10$, here with arbitrary units. The phases of the aforementioned Fourier components are made to vary with three distinct frequencies, producing the complex patterns of coherent and incoherent interference. Note how the $\delta \phi_1$ dominates in an intermediate eccentricity region, after which it is suppressed because of the interaction between the two components of Eq. \ref{eq:dephasing_master} and the presence of a double integral. In the right panel, the Fourier components are instead given Gaussian envelopes, which qualitatively describes the occurrence of a localised resonance along the binary inspiral (see text). The masses used for this plot are $m_1 = m_2=8$ M$_{\odot}$, while the reference eccentricity at 10 Hz is $e_{10 \rm{Hz}}=0.05$. The plots can be rescaled by varying masses and reference eccentricity. These examples reveal how an eccentric source entering a detector band with non negligible eccentricity can showcase a complex dephasing phenomenology.}
\end{figure*}
In the following, we will use Eq. \ref{eq:dephasing_master} to compute some analytic and numerical results in the low eccentricity limit, in order to explore the phenomenology of the different possible dephasing prescriptions due to the first few epicyclical resonances. We focus on three simple classes of Fourier coefficients, though note that the behaviour in general can be more complex. As a reference, we will analyse the relevance of different dephasing components $\delta \phi_n$, given an observation of a inspiralling binary with total mass $M$ and reduced mass $\mu$ that enters the observable band of some detector at the frequency $f_{\rm in}$, and has a reference eccentricity $e_{\rm ref}$ at a given frequency. This excercise serves to build intuition for more realistic dephasing prescriptions analysed in later sections. Additionally, it demonstrates that the typical scalings assumed for dephasing prescriptions of different EE may not be appropriate for eccentric sources.
\subsection{Constant Fourier coefficients}
As a first example, we assume for simplicity that the force coefficients have no dependence on the current orbital elements of the binary, i.e. they are constant:
\begin{align}
    B_{n}^{\rm{R,S}}(a,e) \to  B_{n}^{\rm{R,S}}
\end{align}
We note however, that analytical results calculations can be easily repeated for forces that present a power law scaling in $a$ or $e$. We can now explicitly calculate the dephasing for different Fourier modes. Inserting Eqs. \ref{eq:adot}, \ref{eq:edot} and the eccentricity evolution equation (Eq. \ref{eq:eevol}) in Eq. \ref{eq:dephasing_master} and integrate from the initial frequency to some final frequency. After some amount of algebraic computation and a series expansion in small eccentricities, we obtain the following result for the dephasing:
\begin{align}
\label{eq:resdephconst}
    \delta \phi &\approx \mathcal{K}(f) \times \bigg[ B_{0}^{\rm S} +  \frac{3e_0(f)}{2060}\left(3106 B_{1}^{\rm S} + i1693  B_1^{\rm R} \right) \nonumber \\
   &+ \frac{7e_0(f)^2}{5856} \ \big(14262 B_{0}^{\rm S} - 1691 B_{2}^{\rm S}\nonumber \\ & - i1230 B_2^{\rm R} \big) \bigg] + \mathcal{O}\left( e_0^3\right),
\end{align}
where {once again $e_0(f)$ is the unperturbed eccentricity, which can evolve due to GW emission. Note that we neglected the negligible contribution of the final frequency to the integral}. The coefficient $\mathcal{K}$ reads:
\begin{align}
    \mathcal{K}(f)= \frac{25 c^{10} f^{-14/3}}{32256 \left( G^{11}M^{5}\mu^6 \pi^{17}\right)^{1/3}}.
\end{align}
Eq. \ref{eq:resdephconst} demonstrates how different contributions to the total dephasing of a binary source can be decomposed into the contributions from different Fourier modes of the perturbative force. In the case of constant Fourier Coefficients, the contributions from $n>0$ resonances are suppressed by powers of the eccentricity. Therefore, they scale with an additional factor $f^{-19/18}$ with respect to the scaling expected due to the constant components (see Eq. \ref{eq:eevol}). In general, this implies that they will be more significant at the very early stages of the GW observation. The presence of different scalings is a key aspect in order to extract the different dephasing contributions from a realistic signal \citep[see e.g.][for work on distinguishing multiple environmental effects in a single GW signal]{2023NatAs...7..943C}. The a-priori knowledge that several dephasing contributions may be present in a signal, with scalings related by powers of $f^{-19/18}$, may be used to better extract the several distinct Fourier modes of the force. We also see that, roughly, the order of magnitude of the contributions scale as:
\begin{align}
\label{eq:order_mag}
    \frac{\delta \phi_n}{\delta \phi_0} \sim \rm{few}\times (e_{\rm in})^n \left( \frac{B_n^{\rm S,R}}{B_0^{\rm S,R}}\right),
\end{align}
and for the first few Fourier modes we have e.g.:
\begin{align}
    \delta\phi_0 &= \mathcal{K}(f)\left( 1 + \frac{99827}{5856}e_0(f)^2 + \, ...\right)B_{0}^{\rm S}\\
    \delta\phi_1 &= \mathcal{K}(f) \frac{3e_0(f)}{2060}\left(3104B_{1}^{\rm S} + i 1693B_1^{\rm R}\right)+\,... 
\end{align}
and so on.

With the aid of Eq. \ref{eq:order_mag}, one can easily estimate what Fourier components of the perturbative force will produce the largest dephasing. We note here again that replacing the constant coefficients with simple power laws does not affect the qualitative results. Power law forms for the Fourier coefficients simply modify the overall frequency scalings, and change some numerical pre-factors of the various contributions to the total dephasing.

Even in this simplified case, i.e. when only considering constant Fourier coefficients and resonances up to $n=2$, the phenomenology of possible dephasing behaviors is extremely rich. Figure \ref{fig:phenomenology_constant} shows three possible cases, in which the dephasing of the binary evolves in wildly different ways, depending only on the phases of the Fourier coefficients. For concreteness, we consider here a binary with components $m_1 = m_2=8$ M$_{\odot}$ with a reference eccentricity at 10 Hz of $e_{10 \rm{Hz}}=0.05$, though the results can be rescaled for any mass and reference eccentricity. The binary is perturbed by a normalised force with $\lvert B_{1}^{\rm S} \rvert=1$, $\lvert B_{1}^{\rm S} \rvert=3$ and $\lvert B_{2}^{\rm S} \rvert=15$, here with arbitrary units. The values are chosen to showcase and highlight different behaviours, though are entirely representative (see section \ref{sec:astro_numerical}). In the first plot, the phases Fourier coefficients are chosen such that the total dephasing coherently adds up to a value larger than the naive orbit averaged expectation. In the second plot, the phase of $B_{1}^{\rm T}$ is changed by 180$^{\circ}$. Then, the total dephasing of the binary can change sign twice, transitioning within regimes where $B_0^{\rm S}$, $B_{1}^{\rm S}$ or $B_{2}^{\rm S}$ dominate. In the last plot, the phase of $B_{2}^{\rm S}$ is similarly shifted such that the total dephasing is simply reversed with respect to the expectation of the orbit averaged force.

\subsection{Fourier coefficients with slowly varying phase}

We now analyse the case of Fourier components with an adiabatically changing phase, i.e. components of the form:
\begin{align}
\label{eq:adi}
    B^{\rm S,R}_n = \lvert B^{\rm S,R}_n\rvert \exp \big[ 2 \pi i \lambda _n(a,e) t \big],
\end{align}
where the $\lambda_n$ are additional frequencies that determine the slow variation in the phase of the Fourier components.

We highlight here, that external forces with a fixed frequency will in general transition between being ``fast" with respect to the orbital frequency to being adiabatic as the binary is chirping. They will therefore transitioning through a regime where $\lambda_n\sim n f_{\rm K}$ to  a regime where $\lambda_n <  f_{\rm K}$. In fact, this scenario is what is typically associated with the concept of resonance in orbital dynamics. Such resonances have been shown to be able to produce important dephasing for EMRI systems \citep{2019Bonga,2021speri}, and may potentially leave detectable dephasing imprints in the GW inspiral of neutron star binaries \citep{2024janos}.

The parameterisation shown in Eqs. \ref{eq:perturbingforces} represents the low frequency limit for $\lambda_n$. As an example, such Fourier components could be produced by the gravitational of a third body, that is slowly orbiting around the binary at a fixed frequency, thus changing the orientation of the perturbative force with respect to the binary's argument of pericenter. As an additional example, numerical simulations of accreting binary systems show that features of the gas flow close to the binary are sourced by the larger scale accretion flow, often by the behaviour of the gas at the circumbinary disc cavity edge. Both of these possibilities will be discussed in more detail in later sections. Here we simply analyse the phenomenology of the resulting dephasing prescriptions.

An example of dephasing is shown in the first panel of Fig. \ref{fig:phenomenology_other}, in which we chose $\lvert B_{0}^{\rm S} \rvert = 1$, $\lvert B_{1}^{\rm S} \rvert = 3$ and $\lvert B_{2}^{\rm S} \rvert = 10$. Here however, the phases of the Fourier components are varying with a frequency of $0$, $1$ and $3$ milli Hz, respectively. The resulting interference causes the total dephasing of the binary to vary wildly. We note in particular how the $n=1$ component $\delta \phi_1$, which arises as a consequence of an excitation of the perturbed eccentricity $\delta e$, only dominated in an intermediate region where it is not suppressed by a the double integral implied by Eq. \ref{eq:dephasing_master}.

\subsection{Fourier coefficents with Gaussian envelopes}
\label{sec:sec:localised}
The final important class of Fourier coefficients is the one that represent transient physical forces that only couple to the binary at very specific times, frequencies, or eccentricities. In our framework, we describe the qualitative features of such localised forces by introducing Gaussian envelopes to the Fourier coefficients:
\begin{align}
    \lvert B^{\rm S,R}_n \rvert  \propto  \exp \left( \frac{(a - a_{\rm{ res};n})^2}{\sigma_{a;n}^2} \right)\exp \left( \frac{(e - e_{\rm{ res};n})^2}{\sigma_{e;n}^2} \right),
\end{align}
where the $a_{\rm{res};n}$ and $e_{\rm{res};n}$ denote the orbital elements at which the localised forces can affect the binary. The standard deviations describe instead the sharpness of the localised force, as a function of the orbital elements. Note that here we restrict our showcase to coefficients with constant phase, though in reality they may also vary.

In the second panel of Fig. \ref{fig:phenomenology_other}, we show the resulting dephasing for a choice of Fourier coefficients with a Gaussian envelope. The localised resonances have been chosen to occur at a frequency of 10, 5 and 2.5 Hz for the $n=0$, $n=1$ and $n=2$ components, respectively, and given a standard deviation of $10\%$ of the given frequency at wich the force perturbation peaks. The amplitude of the Fourier components is once again chosen preserving the same ratios as in the previous sections. We see how the dephasing inherits the signatures of the three distinct peaks, appearing similarly to cumulative Gaussian error functions.

\section{Common Analytical prescriptions for environmental effects}
\label{sec:astro}

\begin{table*}
    \centering
    \begin{tabular}{c|c|c|c|c}
      Perturbation & Third body &  Subsonic drag ($q=1$)&  Supersonic drag ($q=1$) & Accretion drag ($q=1$)  \\ \hline \vspace{0.3cm}
       $\xi$ & $Gm_3R_3^{-3}$ &  $G^{5/2}M^{3/2}\rho c_{\rm s}^{-3}$ & $G\rho$ & $G^2M\rho c_{\rm s}^{-2}$ \\

      $B_{0}^{\rm S}/\xi$ & 0 & $-\pi/3 \times(2-e^2)a^{-1/2}$ &  $-4\pi a(4+7e^2)$ & $-4\pi(1 - e^2)$ \\
      $B_{1}^{\rm S}/\xi$ & 0 & $\pi/3\times ea^{-1/2}$ & $16\pi ae$ & 0 \\
      \vspace{0.3cm}
      $B_{2}^{\rm S}/\xi$ & 0 & $-\pi e^2a^{-1/2}/3$ & 0 & $\pi e^2/2$\\ 
      $B_0^{\rm R}/\xi$ & $-a(2+e^2)/2$ & 0 & 0  & 0\\
      $B_1^{\rm R}/\xi$ & $ ae/2$ & $\pi/3\times ea^{-1/2}$ & $8\pi i ae$ & $4\pi e$  \\
      $B_2^{\rm R}/\xi$ & $ae^2/4$ & $i\pi e^2a^{-1/2}/3$ & $-4\pi i ae^2$ & $2\pi e^2$

    \end{tabular}
    
    \caption{Fourier coefficents for the analytical EE prescriptions analysed in this work.}
    \label{tab:coeff}
\end{table*}
We now analyse how even simple models of astrophysical perturbations on eccentric inspiralling binaries produce time varying forces with various Fourier components. In particular, we analyse the effect of tidal forces due to a third massive body, as well as various prescriptions for hydrodynamical drag. These EE are key in determining the evolution and merger efficiency of binaries in the dynamical \citep{antonini2016b,antonini2017,silsbee2017,toonen2018,rodriguez2018,vignagomez2021,martinez2020,arcasedda2021,trani2022} and AGN channel \citep{Antoni:2019, LiLai:2022, Dempsey3D:2022, rowan2023,2023whitehead, DittmannDempsey:2024,dittmann2024,Rowan2024_rates,2022liu}, and are also commonly used to describe perturbed EMRI systems \citep{2022coogan,2022cole,2024dyson,2025dyson}. These insights from simple analytical models will be compared to the results of various numerical approaches in section \ref{sec:astro_numerical}.

\subsection{Three body systems}
We analyse the gravitational perturbation caused by a third body of mass $m_3$ on the binary system. For simplicity, we place the third body in the orbital plane and at a fixed position. The perturbative gravitational force on the binary is the difference of the gravitational force on each component:
\begin{align}
    {\rm{d \mathbf{F}}} = \frac{G m_{3}}{R_{13}(t)^2} \mathbf{n}_{13}(t) - \frac{G m_{3}}{R_{23}(t)^2} \mathbf{n}_{23}(t)
\end{align}
where $R_{ij}$ the separation between either binary component and the third body, while and $\mathbf{n}_{ij}$ is the corresponding unit vector. Here we perform a Fourier analysis in the limit $R_{i3} >> a$ and to second order in eccentricity. The first step is to decompose the perturbing force ${\rm{d}}\mathbf{F}$ in its radial and azimuthal components:
\begin{align}
    R &= {\rm{d \mathbf{F}}} \cdot \mathbf{n}_{12}, \\
    S&= {\rm{d \mathbf{F}}} \cdot\frac{ \mathbf{v}_{\perp}}{v_{\perp}},
\end{align}
where $\mathbf{v}_{\perp}$ is the velocity component perpendicular to the binary separation vector. Then, a Fourier analysis can be performed by simply expanding all expressions in terms of the epicycles as in Eq. \ref{eq:centers}, where the resulting Fourier coefficients can be read off from the formulas. They are reported in Table \ref{tab:coeff}, and can be used to derive the evolution equations and the resulting dephasing prescriptions via Eqs. \ref{eq:adot} and \ref{eq:edot}.

As expected, the magnitude of the Fourier coefficients scales as a tidal perturbation i.e. with $\xi_{\rm 3b}/a= G m_3  R_3^{-3}$. {We also find that no azimuthal component exists, due to the symmetry of the tidal force expanded to linear order}. Therefore, the magnitude of the dephasing induced by the perturbing forces is entirely dominated by the eccentricity evolution, which is determined by resonances with the first few epicycles, rather than a drift due to an orbit averaged force. Note that, even though radial forces do not induce a direct drift in the orbital elements, they can still result in indirect dephasing of the GW signal due to the evolution of a binary in a modified conservative potential. These types of dephasing are typically analysed in for EMRI systems in dark matter halos \citep{2022speeney}, and their effect for equal mass binaries is discussed in \cite{2023zwick,2025zwick}.

\subsection{Gas embedded binaries}
We now repeat similar calculations for the case of binaries perturbed by hydrodynamical forces. As discussed previously these can affect EMRIs in the wet formation channel, massive BH binaries in circumbinary discs and stellar mass binaries in the AGN channel. We consider three types of force, motivated by the many studies of dynamical friction on embedded binary systems. Firstly, we separately consider the subsonic and supersonic limit of gaseous dynamical friction, as derived by \cite{ostriker1999}. Then, we also consider the subsonic limit of Bondi-Hoyle-Littleton drag \citep{bondi1952,fabj2020,Donmez:2012cq,Beckmann2018}, which captures the dissipative effect of accretion. The perturbative forces on the individual components are:
\begin{align}
    {\rm{d}}\mathbf{F}_i^{\rm sub} &\approx \frac{4 \pi G^2m_i^2}{v_i^2}\rho\times\frac{\mathbf{v}_i}{v_i} \times\frac{v_i^3}{3 c_{\rm S}^3} \\
     {\rm{d}}\mathbf{F}_i^{\rm sup} &\approx \frac{4 \pi G^2m_i^2}{v_i^2}\rho\times\frac{\mathbf{v}_i}{v_i}\times \ln(\Lambda) \\
     {\rm{d}}\mathbf{F}_i^{\rm BHL} &\approx \frac{4 \pi G^2m_i^2}{c_{\rm s}^2}\rho\times\frac{\mathbf{v}_i}{v_i}
\end{align}
where $\rho$ is the gas density, $c_{\rm s}$ the speed of sound and the superscripts refer to the subsonic, supersonic and BHL gas drag limits. The perturbing force is then computed as ${\rm{d}}\mathbf{F} =  {\rm{d}}\mathbf{F}_1 -  {\rm{d}}\mathbf{F}_2$. The Fourier analysis of both radial and azimuthal force components can be performed in a manner similar to the previous section, and the resulting Fourier components are reported in Table \ref{tab:coeff}.
The magnitude of the various Fourier coefficients for hydrodynamical forces scales with different combinations of the binary and gas parameters. In general, we find non-zero constant components in the azimuthal direction, representing the orbit averaged drag experienced by a binary. Both radial and azimuthal oscillations in the force are excited by the presence of eccentricity. We find that for all analytical force prescriptions:
\begin{align}
    \lvert B_n\rvert \propto e^n,
\end{align}
a simple scaling which we will use to compare with numerical results. The magnitude of the $n=1$ and  $n=2$ components is sub-dominant for these analytical prescriptions. However, we will see that this conclusion fails to capture the reality of gas-embedded binaries.

\begin{figure*}
    \centering
    \includegraphics[width=0.33\linewidth]{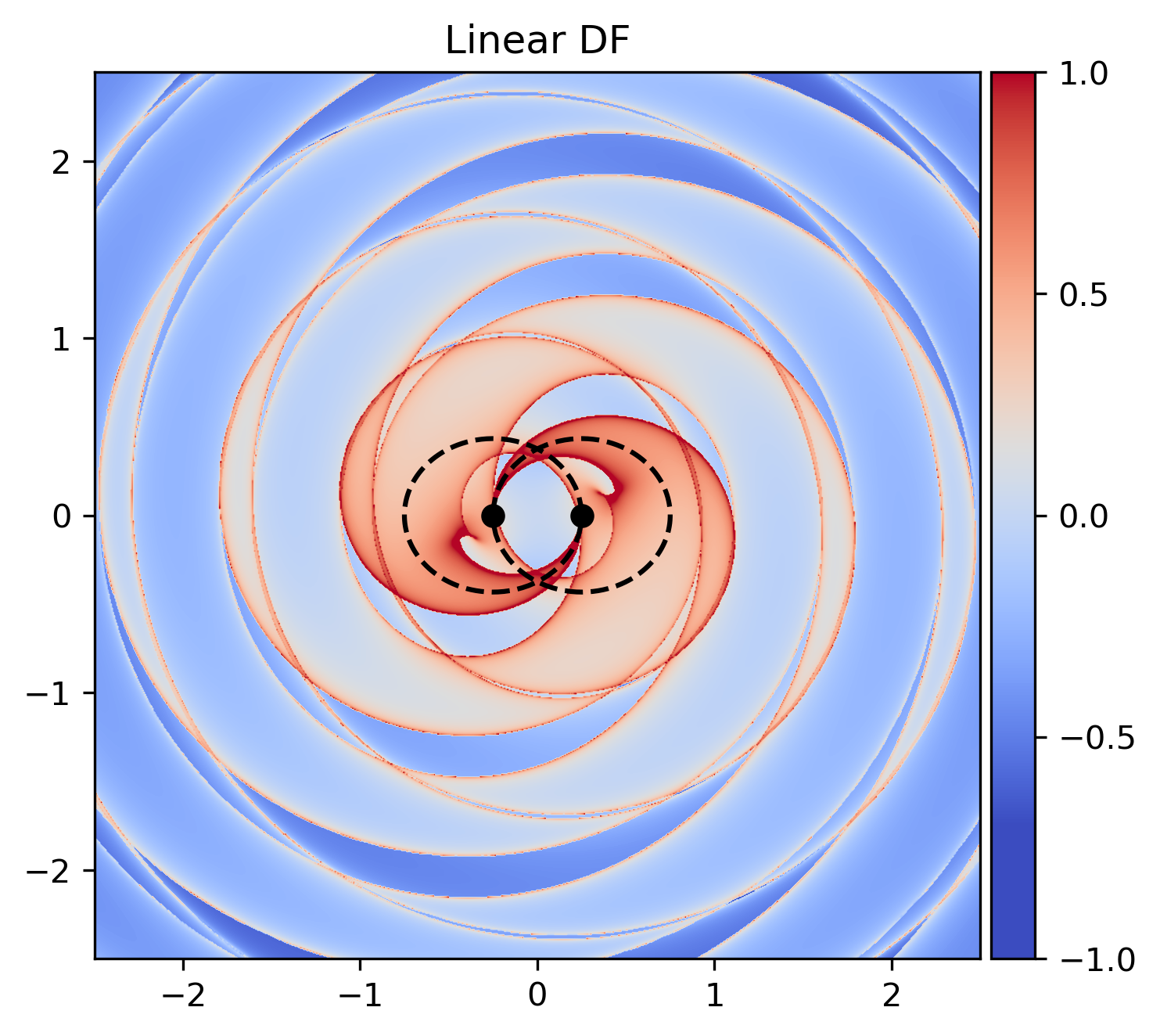}\includegraphics[width=0.33\linewidth]{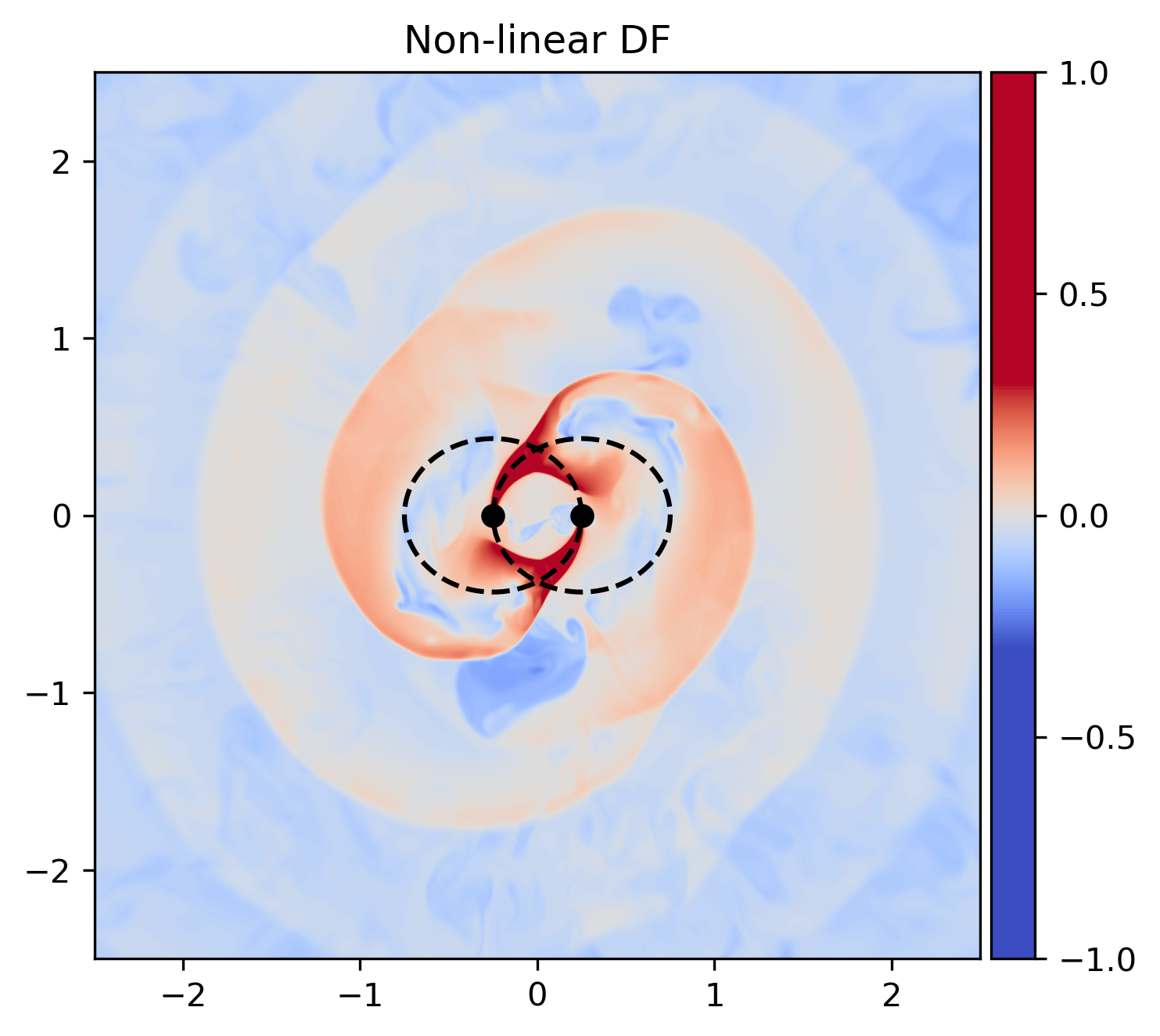}\includegraphics[width=0.33\linewidth]{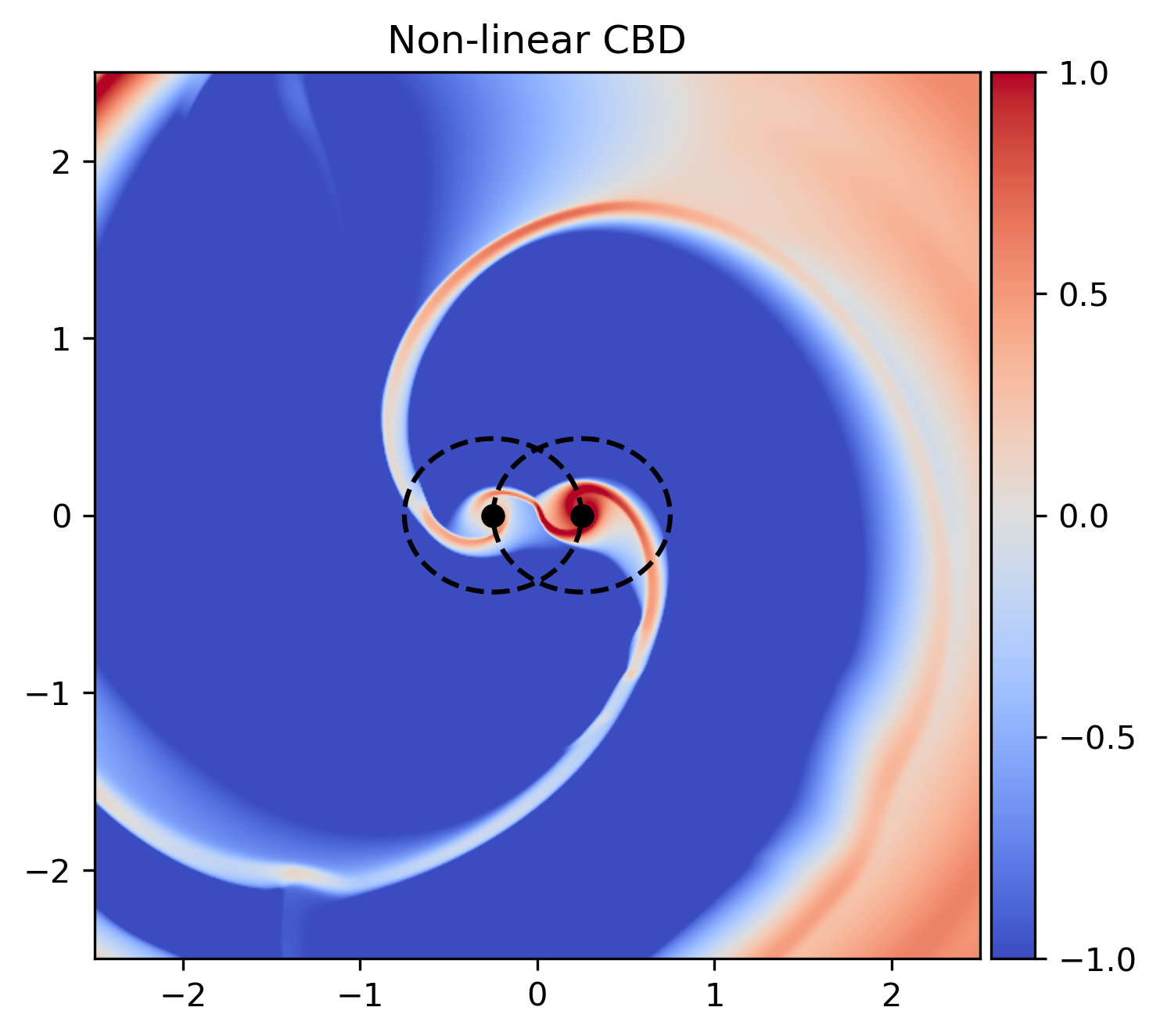}
    \caption{Gas density snapshots showcasing the three approaches detailed in section \ref{sec:astro_numerical}, i.e. dynamical friction in the linear regime (left panel - see section \ref{sec:david}) and non-linear regime (middle panel - see section \ref{sec:philip}), as well as a fully non-linear CBD simulation (right panel-see section \ref{sec:chris}). The reference eccentricity here is chosen to be $e=0.5$. The Mach number for the first two panels is 8, here referring to the velocity of the orbit at pericenter passage with respect tot the speed of sound. The Mach number for the CBD simulation is $\sim 10$, here referring to the aspect ratio of the CBD disc. The different gas morphhologies are reflected in the forces experienced by the embedded binary, which gives rise to different Fourier coefficients and dephasing. The density is normalised and plotted in logarithmic scale, such red denotes the over densities and blue the under densities.}
    \label{fig:morphology}
\end{figure*}
\section{EE in gas embedded binaries: Numerical experiments}
\label{sec:astro_numerical}

In the following, we analyse the consequences of the fact that gas embedded binaries in realistic astrophysical environments, be it circumbinary discs (CBD) or AGN discs, are subject to perturbative forces dominated by variability and fluctuations. We start by briefly reviewing existing literature on the topic, before presenting a curated sample of numerical experiments with three different approaches. We start with a semi-analytical treatments of gas dynamical friction in the linear regime (\ref{sec:david}), and investigate the transition to the non-linear hydrodynamics with a few numerical experiments (\ref{sec:philip}). Finally, we discuss the implications of a fully non-linear simulation of a CBD disc (\ref{sec:chris}). These experiments demonstrate how adding more realistic treatment of hydrodynamics inevitably leads to strong variability in the forces, and that in many cases the resulting dephasing is dominated by force Fourier modes higher than $n=0$.
\begin{figure*}
    \centering
    \includegraphics[width=0.95
    \columnwidth]{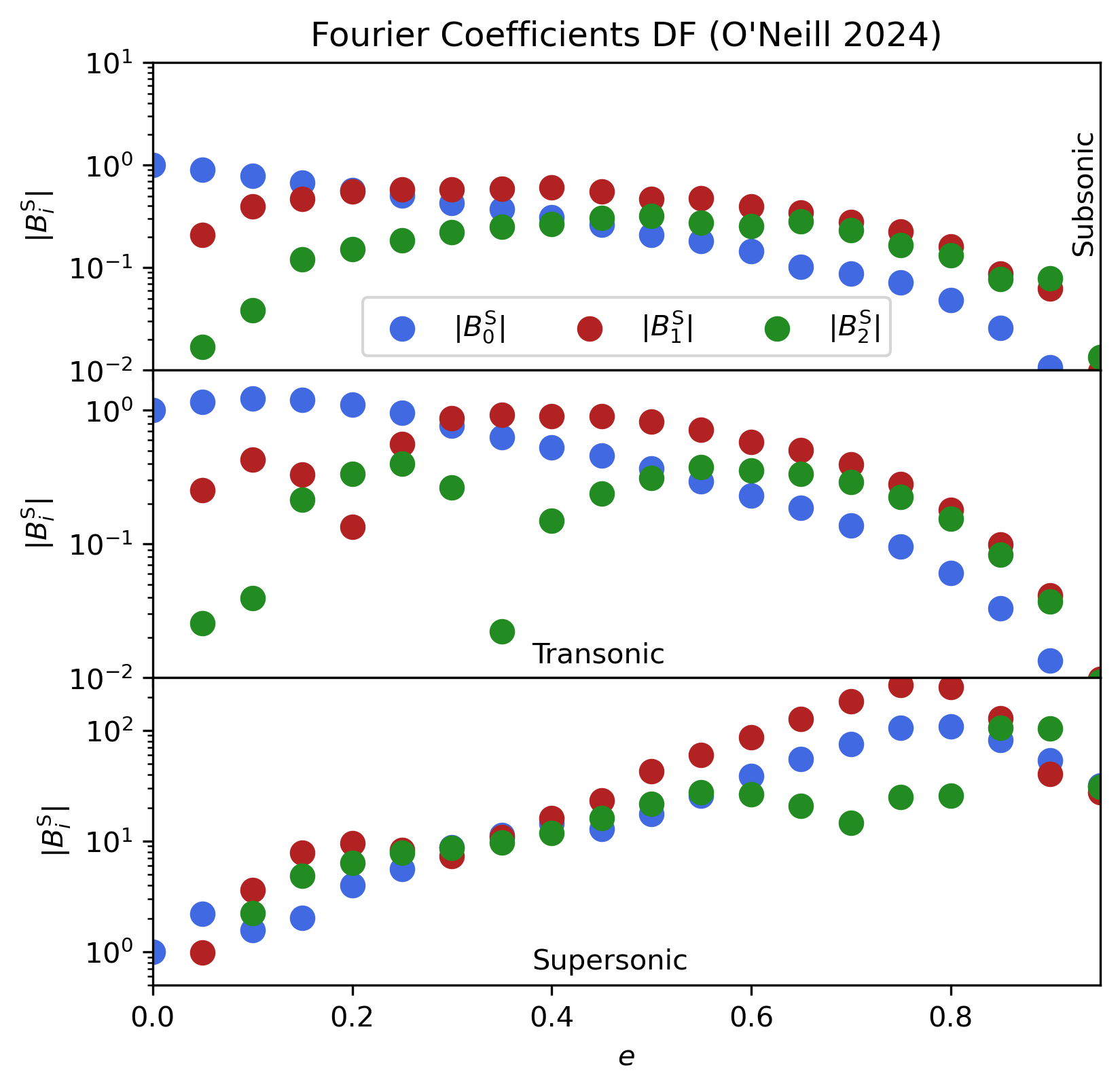} \includegraphics[width=0.95
    \columnwidth]{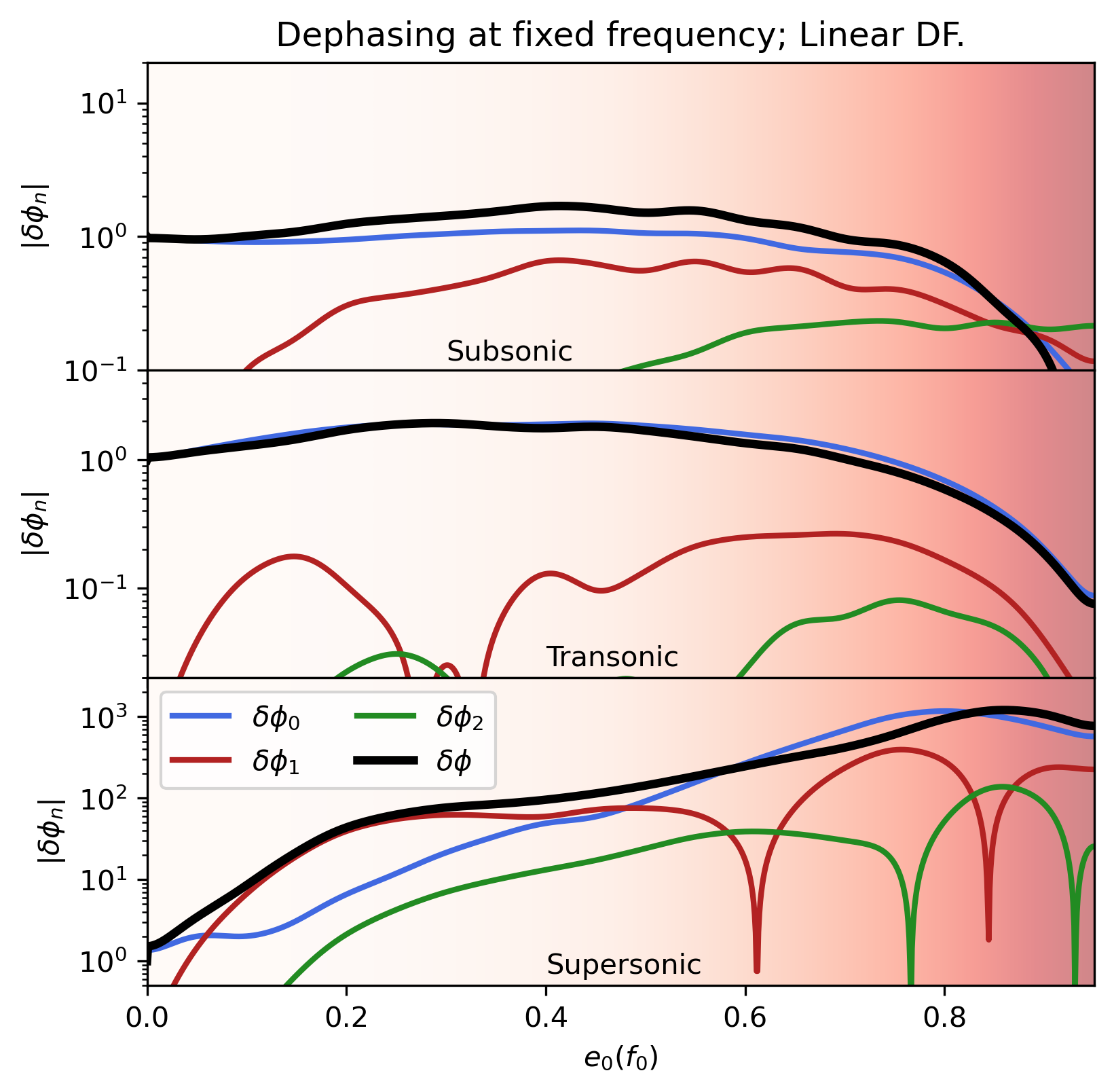}
    \caption{Left panel: Fourier coefficients (here only for azimuthal forces) for gaseous dynamical friction in the linear response regime, as a function of eccentricity {for binaries at a reference frequency $f_0$} (see section \ref{sec:david}). The subsonic, transonic and supersonic cases correspond to a mach number at peri-apsis of 0.5,1.5 and 8, respectively. In the supersonic case, the dephasing caused by the $n=1$ resonance dominates in the range  $e_0\sim 0.05$ to $e_0 \sim 0.4$. Right panel: The total dephasing $\delta \phi$ experienced by the binary {for the given eccentricity at the reference frequency}, as well as its components $\delta \phi _n$. Note how there are vast regions in which the dephasing caused by the $n=1$ and $n=2$ Fourier modes dominates over the $n=0$ component. The dephasing curves are computed by means of Eq. \ref{eq:resdephconst}, meaning that the high eccentricity results are only illustrative (represented here by the shading).}
    \label{fig:oneill}
\end{figure*}
\begin{figure*}
    \centering
    \includegraphics[width=0.95
    \columnwidth]{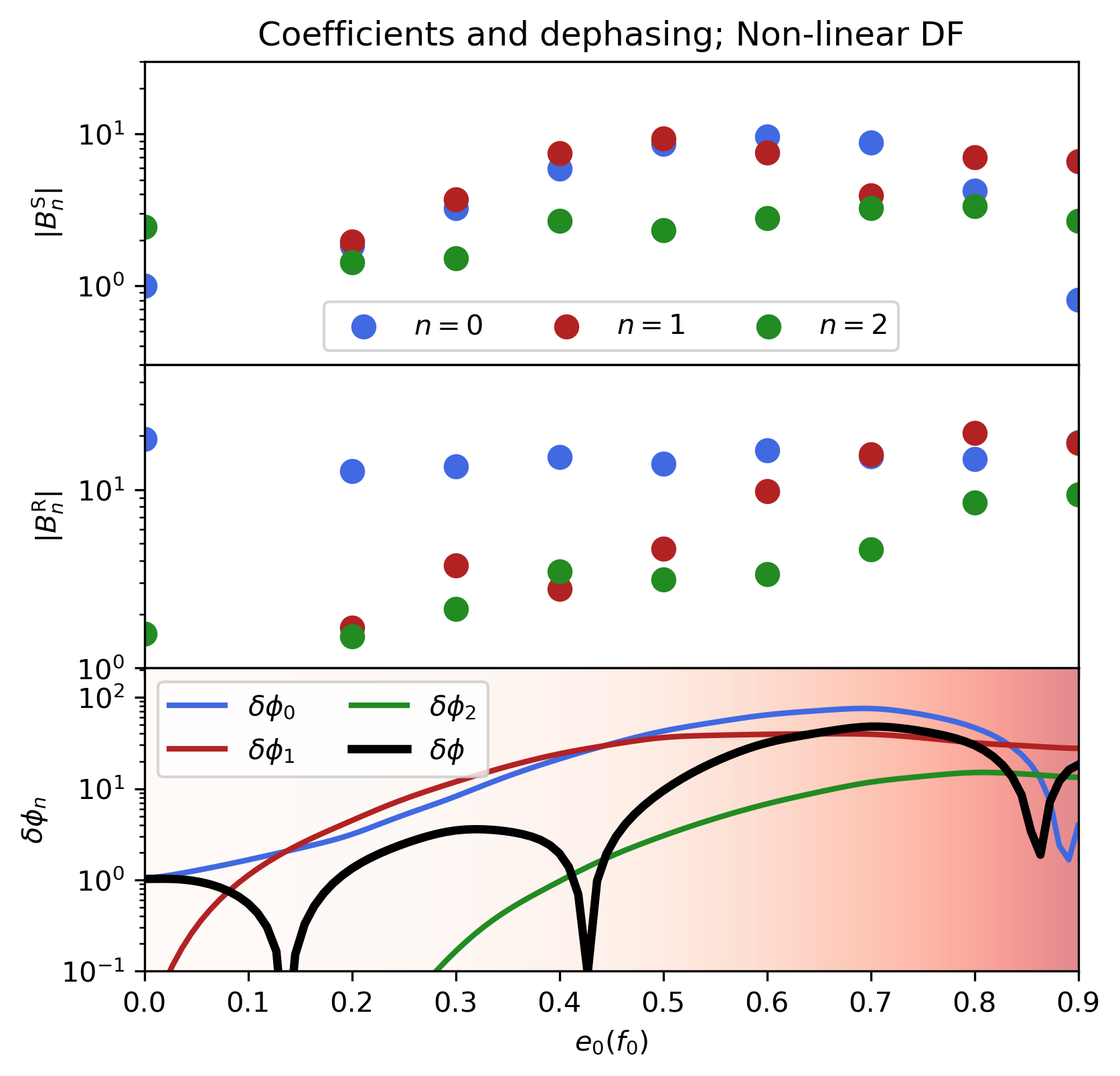}
    \includegraphics[width=0.95
    \columnwidth]{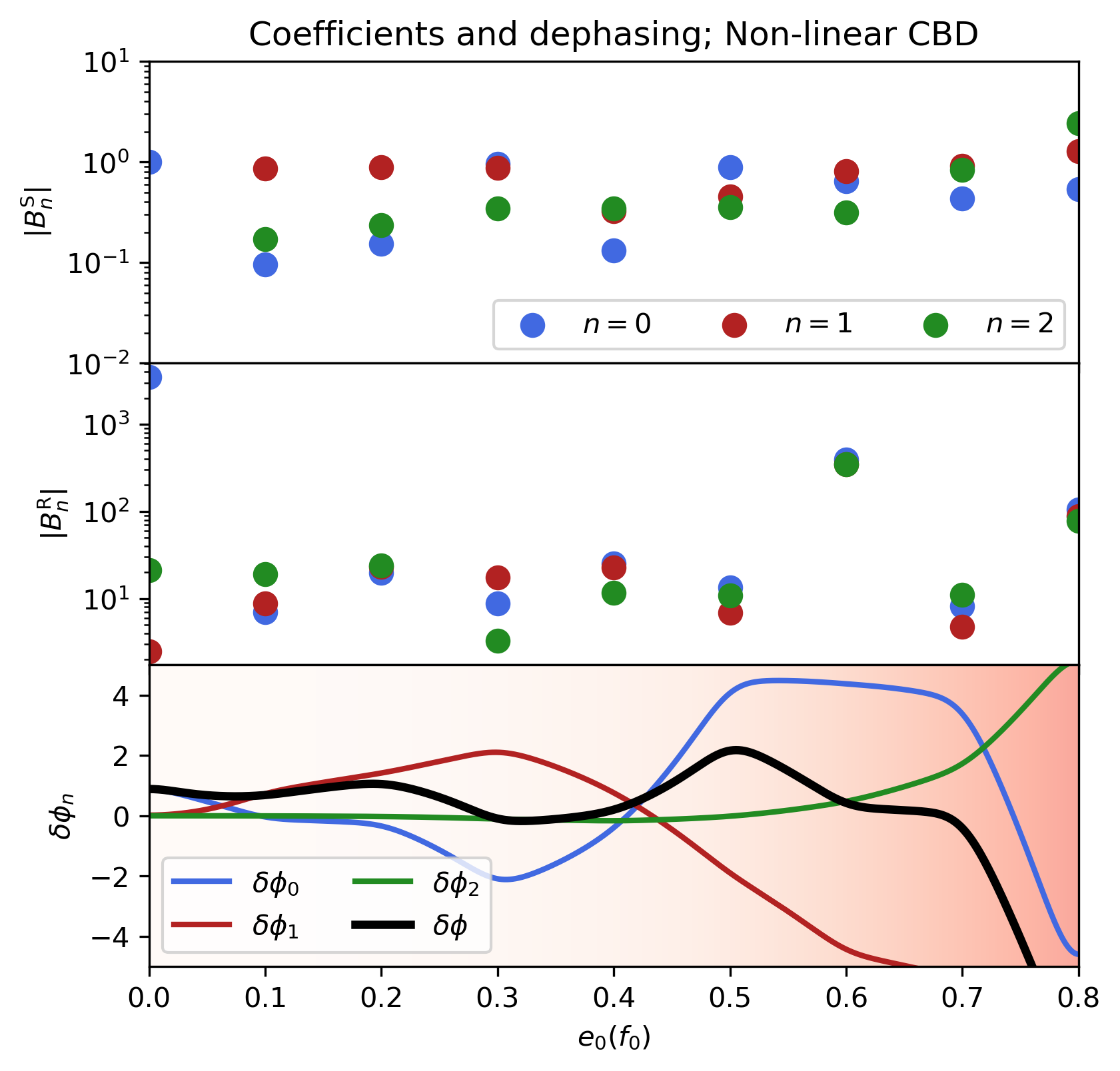}
    \caption{Left panel: Fourier coefficients for the forces experienced by an equal mass binary at a given reference frequency subject to gas dynamical friction in the non-linear regime (see section \ref{sec:philip}). The resulting total dephasing {for the given eccentricity at the given reference frequency} is plotted along with its components $\delta\phi_n$ in the bottom panel, and is dominated by the $n=1$ resonance between $e_0\sim 0.1$ and $e_0\sim 0.3$. Right panel: Fourier coefficients for the forces experienced by an equal mass binary embedded in a CBD (see section \ref{sec:chris}). The resulting total dephasing at a fixed frequency is plotted along with its components $\delta\phi_n$ in the bottom panel, and is dominated by the $n=1$ resonance between $e_0\sim 0.03$ and $e_0\sim 0.2$. {The value of the dephasing is normalised, such that only the relative size matters} Note how the total dephasing switches between negative to positive at an eccentricity of $\sim 0.4$. We plot this specific curve not in logarithmic scale to highlight the resemblance the well known eccentricity attractor in CBD at $e\sim 0.4$ \citep{2021doraziodisc,Franchini:2022}. Here we observe that the switch is a result of the interference in the evolution of the binary between the $n=0$  and the $n=1$ Fourier modes. The dephasing curves are computed by means of Eq. \ref{eq:resdephconst}, meaning that the high eccentricity results are only illustrative (represented here by the shading).}
    \label{fig:chris}
\end{figure*}
\subsection{Insights from existing hydrodynamical simulations}
Massive \citep{DOrazioCharisi:2023,D'Orazio:binlite:2024} and stellar mass \citep{2024whitehead,rowan2023} BH binaries embedded in gaseous circumbinary discs (CBD) are capable of producing smoking gun electromagnetic signatures, and are thought to constitute an important fraction of GW mergers both for space based and ground based GW detectors. Additionally, gas embedded EMRIs are thought to be an important contributor to the expected GW detection rate, and may be related to the recently discovered class of electromagnetic quasi-periodic eruptions \citep{2010Dai,2023franchini,2023Linial}. Thus, due to their enormous relevance for both GW astrophysics and conventional electromagnetic observations, the dynamics of gas embedded binaries are the subject of numerous studies. Here we briefly review what such studies suggest regarding variability.

Massive BH binaries are thought to arise from galaxy mergers, which efficiently channel material into the nucleus of the newly formed galaxy \citep{Barnes_Hernquist_1996, Springel:2005}. As a result, they are likely to undergo phases of gas driven evolution that play a key role in bridging the gap from wide separations to GW-driven inspirals \citep{Begel:Blan:Rees:1980, Milosavljevic:2003, 2009haiman}. Recent research further indicates that gas surrounding the binary can remain dynamically coupled to the system even deep into the GW-dominated regime \citep{Farris_dec+2015, Dittmann:Decoupling:2023, Krauth:2023, Avara:3DMHD:2023, Cocchiararo:2024, ONeillTiedeDOrazio:2025}, and that LISA may be sensitive to secular gas-driven modifications to the binary’s orbit \citep{garg2022, 2024garg, 2023Tiede}. A major focus of numerical studies has been the periodic nature of the accretion rate onto the binary \citep[see][for a recent review]{LaiMunoz:Review:2022} and its implications for identifying massive compact binaries through variability in electromagnetic signatures \citep[e.g.][]{D'Orazio:binlite:2024,2024franchini}. On the contrary, most recent work on gas-induced orbital evolution has emphasized the secular, time-averaged effects, with limited attention paid to their time-dependent behavior. Nevertheless, a few works -in particular \cite{Roedig_Trqs+2012}- have conducted a spectral analysis of the torques in these systems. The computed torque power spectra showcase clear dominant peaks at small multiples of the orbital frequency that dominate over the constant component, sometimes by a factor of $\sim 100$. More recently in \cite{2024zwick}, the torque spectrum of a single simulation of a circular embedded massive BH binary was analysed in the context of GW observables. It was shown that the torques exhibit significant variability, displaying features beyond those associated solely with the binary’s orbital frequency. Notably, the torque magnitude fluctuates by a factor of approximately 30 relative to its orbit-averaged value. We expect similar considerations to also apply to CBDs around stellar mass binaries, which are expected in the AGN channel.

We now turn our attention to BHs embedded in larger scale accretion discs, i.e. EMRI systems. As discussed thoroughly in \cite{2022zwick}, torque fluctuations in gas-embedded systems can arise as a result of several physical mechanisms. Although linear torque theory \cite{GoldreichTremain:1980} assumes laminar gas flows, real accretion discs are expected to be turbulent due to the magneto-rotational instability \citep{1991ApJ...376..214B,secunda18}. A migrating perturber traveling through such a turbulent medium will encounter density fluctuations, leading to stochastic variations in the experienced torque. Numerous studies in the context of planetary migration confirm that turbulence introduces a random component to the torque, superimposed on the net, linear (Type I) migration \citep{2005nelson,2024WuChenLin}. When the perturber-to-disc mass ratio is small, these stochastic fluctuations can dominate the migration process entirely. Intrinsic non-linearity in the gas flow near a sufficiently massive secondary object can also induce torque variability. This has been observed in high-resolution simulations of intermediate mass-ratio inspirals ($q = 10^{-3}$) embedded in two-dimensional, laminar discs \citep{2021andrea}. The latter have also been analysed in \cite{2024zwick}, showing that fluctuations at a few multiples of the orbital frequency exceed the orbit averaged value by factors of several hundreds. Importantly, both the amplitude and frequency of these variations grow with increasing Mach number and simulation resolution, suggesting a genuine physical origin potentially linked to large, super-orbital torque oscillations \citep{2021andrea}. Finally, several physical processes could drive such variability deep within the secondary’s influence radius. These include stochastic accretion events \citep{Kelly:2011ku}, magnetohydrodynamic effects, turbulent drag \citep{2024WuChenLin}, or other small-scale gas dynamics.

Overall, the emerging picture is that the intrinsic non-linearity of the hydrodynamics of embedded binaries necessarily induces strong variability in the resulting forces. In the presence of eccentricity, it is therefore likely that the ultimate driver of any perturbation to the binary's orbital elements is a resonant effect as modeled by Eq. \ref{eq:adot}. It is also worth noting that most numerical studies on this topic—have primarily focused on discs with a characteristic scale height of $h/r = 0.1$. Steady-state models indicate that discs surrounding such massive central objects are expected to be significantly thinner, with $h/r \sim \mathcal{O}(10^{-2} - 10^{-3})$ \citep{AccretionPower, sirko2003, thompson2005}. A growing body of recent work has demonstrated that both the binary's dynamical evolution and the morphology of accretion flows are sensitive to the disc's scale height \citep{tiede2020, Dittmann:2022, Penzlin:2022, Franchini:2022}. In particular, the total gravitational torque exerted on the binary—as well as the amplitude of its fluctuations—has been shown to increase significantly as $h/r$ decreases \citep{tiede2020, 2021andrea}.

\subsection{Numerical approaches}
Here we build up on our insights from analytical models and the existing literature with a succession of numerical experiments. The goal is to build up from a simplified treatment of gas dynamical friction in the linear response regime, to the emergence of non-linear behaviour when the perturber mass is increased, to fully non-linear hydrodynamics. We will see how increasing the complexity of the treatment of the hydrodynamics most often increases the dominance of time-variability. Note that what we provide here is a qualitative comparison rather than a quantitative model. All our experiments are still based on a simplified isothermal treatment of the gas and Newtonian orbital dynamics. In this way, the problem is scale-free and we can expect the results to apply to equal mass binaries of any size. We also limit our analysis to binaries with equal mass ratio, which is likely a conservative case (as EMRIs are more susceptible to high frequency fluctuating forces \cite{2019andrea,2021andrea,2024zwick}). Finally, we assume for simplicity the functional form of Eq. \ref{eq:resdephconst} for the dephasing as a proxy for the full expansion in eccentricity. The setup is as follows:
\begin{itemize}
    \item We compare three sets of semi-analytical and numerical simulations of an equal mass binary embedded in a gaseous medium with varying eccentricity.

    \item We extract the steady-state radial and azimuthal forces experienced by the binary from the simulations.

    \item We perform a Fourier analysis of the steady-state forces and extract the $n=0$, $n=1$ and $n=2$ coefficients as a function of the binary eccentricity.

    \item We compute the resulting dephasing $\delta \phi_n$ caused by the different resonant modes $n=(0,1,2)$ at a fixed reference frequency $f_0$, as a function of eccentricity. The graphs shown here are interpolated and smoothed due to the availability of the simulations, though this does not affect the qualitative results.
\end{itemize}

The differences in the three numerical setups are explained below.

\subsubsection{Gas dynamical friction in the linear regime}
\label{sec:david}
We first consider the case of a low-mass binary in motion through a homogeneous gas medium, i.e, 
gaseous dynamical friction \citep{ostriker1999} for a binary system. Here we closely follow the results presented in \citep{ONeill2024}, which analyse both the gas density perturbations and back-reaction forces onto the binary components, for binaries on prescribed Keplerian orbits. An example of the resulting gas morphology is shown in Fig.~\ref{fig:morphology} for an eccentricity of 0.5 and a Mach of 8, here defined as the ratio of pericenter velocity and gas sound speed. The forces applied to the binary components are then computed by integrating over the gas density perturbation, and decomposing between radial and azimuthal components. 
We calculate the force Fourier components across a wide range of orbital eccentricities and three representative pericenter Mach numbers of 0.5, 1.5 and 8, selected here to represent the sub-sonic, trans-sonic and supersonic regimes (see Fig. \ref{fig:oneill}). Note that BH binaries are known to be far from the linear response regime, and the limitations of this linearised approach are described in detail in \cite{ONeill2024}. Here we use these results as a baseline from which to compare the non-linear approaches. Additionally, we stress that most work on EE simply applies the orbit averaged expectation of dynamical friction, meaning that even this simplified approach is already more sophisticated than what is typically adopted.

\subsubsection{Gas dynamical friction in the non-linear regime}
\label{sec:philip}
To illustrate the increase in complexity introduced by the non-linear response to the embedded binary, we perform a suite of 3D inviscid hydrodynamical simulations with Athena++ \citep{Stone2020}. The binary follows a fixed eccentric orbit with a Mach number of 8, here denoting the ratio of the pericenter velocity and the gas speed of sound. Each component modeled as a non-accreting softened point mass potential of softening radius $r_s$. The simulations are performed on a cartesian logarithmic grid with mesh-spacing of $0.018a$ at the $\lvert x\lvert=a$, as was found appropriate by \cite{Kim_2010} for circular binaries. As introduced by \cite{Kim_2009}, the level of non-linearity is measured by $GM/(c_s^2r_s)$, which we set to a representative value of 10. This results in a value of $\mathcal{B}=0.5$ for our chosen softening of $r_s=0.05a$. This value ensures that the non-linear response of the gas will be dominated by the effect of the increased gravitational influence of each component, rather than the curvature of the orbit. However, the non-linearity is still mild enough that no feature similar to a cavity starts to form around the binary. From the resulting density distribution, as seen in the middle panel of Fig. \ref{fig:morphology}, we obtain the timeseries of the radial and azimuthal forces, from which the Fourier coefficients are computed. Though all simulations are run till they reach a quasi-steady state, some still present strong variability of the force components and resulting Fourier coefficients on super-orbital timescales. We obtain some representative Fourier coefficients by performing an orbit-average over the last 20 to 50 orbits and store the standard deviation as a measure of the variability. The slow variability of the coefficients will not be analysed here in detail, as the results serve primarily as a qualitative comparison.

\subsubsection{Fully non-linear CBD simulation}
\label{sec:chris}

Lastly, we consider the time-varying forces on an equal-mass binary on a fixed Keplerian orbit surrounded by a thin, Keplerian accretion disk.
We numerically solve the isothermal Navier-Stokes equations in two dimensions with \texttt{Sailfish} \cite{Sailfish:2024} for nine binary eccentricities ranging from $e \in [0,0.8]$ at a global grid resolution of $1\%$ of the binary semi-major axis.
The physical setup, initial conditions, and force computations are precisely those reported in \cite{SB-CodeComp:2024} for a disk of vertical aspect ratio $h = 0.1$, local sound speed $c_s(r) = h\, v_{\rm kep}(r)$ and constant kinematic viscosity $\nu = 10^{-3} a^2 \Omega_b$ (with the exception of allowing the binary to be eccentric; see also \cite{Zrake2021}). We evolve each system for 2000 binary orbits and output the components of the net force on the binary 50 times per orbit.
The Fourier coefficients are then computed over only the last 1000 orbits after the inner solution has viscously relaxed, and are averaged. Once again, a gas density snapshot is shown in Fig. \ref{fig:morphology}, showcasing the full emergence of a CBD cavity, streams and mini-discs.

\subsection{Results of the numerical experiments}
\begin{figure}
    \centering
    \includegraphics[width=1\columnwidth]{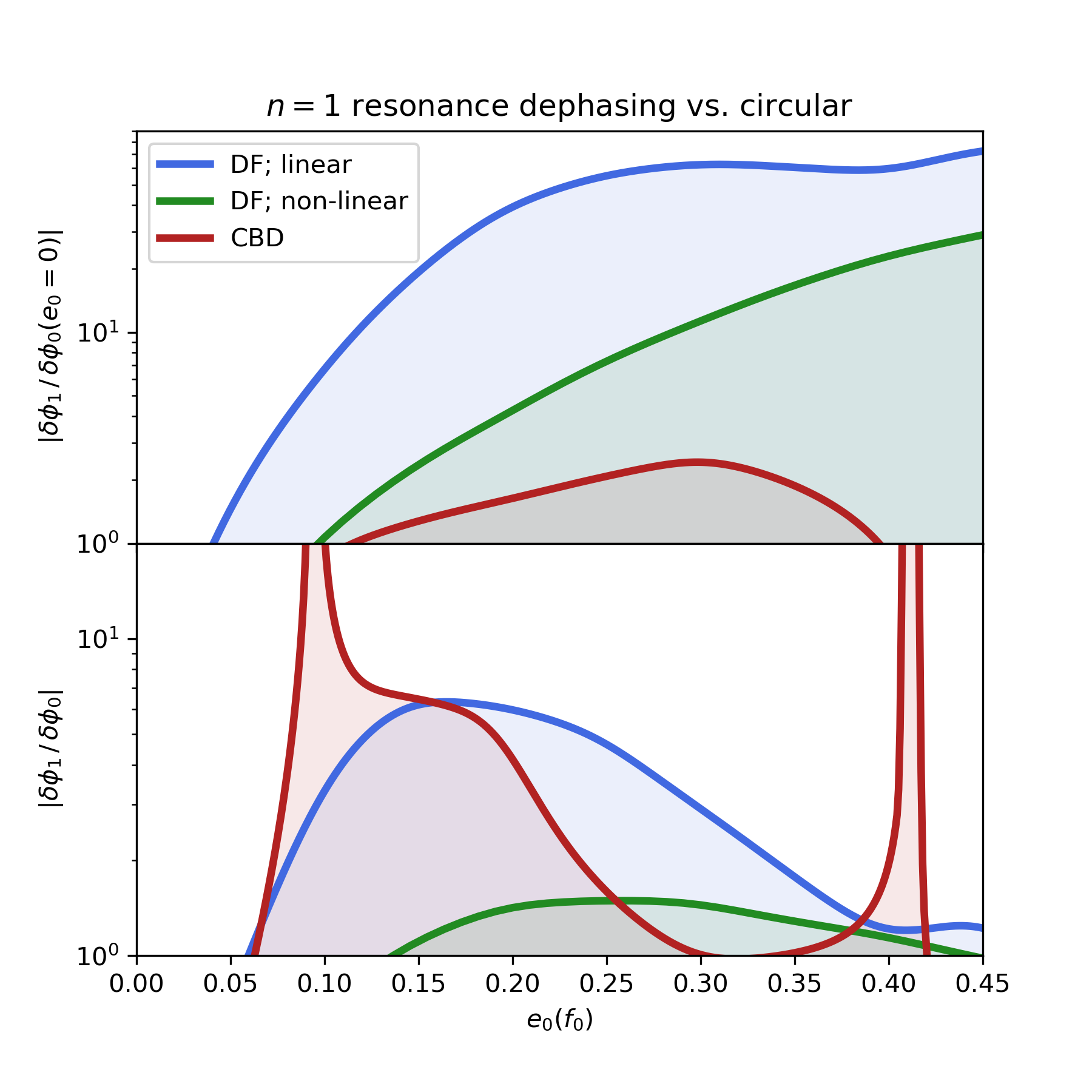}
    \caption{Comparison of the magnitude of the dephasing {for the given eccentricity at a reference frequency $f_0$} caused by the $n=1$ resonance vs the $n=0$ constant component, as a function of eccentricity and for our three reference setups discussed in sections \ref{sec:david} (for a Mach number of 9), \ref{sec:philip} and \ref{sec:chris}. {The value of the dephasing is normalised, such that only the relative size matters}. In the top panel, the $n=1$ dephasing is compared to the expectations for circular orbits. In the bottom panel, it is compared to the dephasing resulting from the $n=0$ constant force component. The plot, along with the results of section \ref{sec:astro_numerical}, strongly indicate that dephasing due to resonances will dominate in gas embedded binaries that retain a modest eccentricity when they enter the band of a GW detector.}
    \label{fig:compare}
\end{figure}
The qualitative results of our investigations can be inferred at a glance by observing the different gas morphologies in the linear, non-linear and CBD regime in Fig. \ref{fig:morphology}. We start by 
describing the linear dynamical friction results, for which the force Fourier coefficients  are shown in Fig. \ref{fig:oneill}. We analyse three different Mach numbers of 0.5, 1.5 and 8, though note that eccentric relativistic binaries are most likely in the supersonic regime (see \cite{2024duque} for an interesting analysis of an EMRI system affected by dynamical friction in the trans-sonic regime). The main take-away is that in all cases the $n=1$ and even $n=2$ modes can dominate for a large range of eccentricities. The resulting total dephasing and the various components $\delta \phi_n$, computed via Eq. \ref{eq:resdephconst}, are shown in the right panel of Fig. \ref{fig:oneill}. Note how in the super-sonic case, the $n=1$ mode determines the total dephasing starting from an eccentricity of $e\sim 0.05$.

The left panel of Fig. \ref{fig:chris} shows both the Fourier coefficients and the resulting dephasing for a set of hydrodynamical simulations at the transition between the linear and non-linear regimes (see section \ref{sec:philip}). As we can see, much of the behaviour of the Fourier coefficients and the dephasing is preserved from the linear case and the $n=1$ mode dominates for eccentricities above $e\sim 0.05$. Interestingly however, the phase of the $n=1$ mode is reversed with respect to the linear case, and the total dephasing behaves differently as a result. At this level of analysis, we do not try to investigate this in detail. We only note that this showcases the importance of accurately modeling the details of force variability when attempting to describe EE in gas embedded binaries. Aspects that are often neglected, such as the phase of the force Fourier modes, can determine wether energy is added to or extracted from the binary.

The right panel of Fig. \ref{fig:chris} shows the Fourier components and the resulting dephasing from a simulation of a binary in a CBD. The $n=1$ and $n=2$ components of the forces dominate for a large range of eccentricities, reflecting the strong variability analysed in e.g. \cite{Roedig_Trqs+2012}. Simulations such as the ones described in section \ref{sec:chris} represent the state of the art in modeling equal mass binaries in thin discs, and the results presented here show that for eccentricities of $\sim 0.05$ and above, the dephasing due to the presence of gas will be dominated by the amplitude of the variability rather than orbit averaged quantities. In fact, this transition eccentricity of $e\sim 0.05$ seems to be quite robustly preserved despite the vastly different numerical approaches whenever the binary are orbiting in the supersonic regime.

We summarise the results from the various numerical approaches in Fig. \ref{fig:compare}, which shows the magnitude of the dephasing caused by the $n=1$ resonance with respect to the $n=0$ component. Once again, we find that in all cases, dephasing will be dominated by the first resonance fora a significant range of eccentricities, typically between $e\sim 0.05$ and $e\sim 0.4$, though at the higher limit our approximations start breaking down. The consequences of this fact for the observational prospects of gas induced EE in eccentric binaries are twofold:
\begin{itemize}
    \item Typical power law scalings used for gas induced dephasing must, at a minimum, be supplied with additional eccentricity scalings, as in Eq. \ref{eq:resdephconst}.

    \item The magnitude of the dephasing will be determined by combinations of gas parameters that relate to force variability rather than average magnitude. As an example, a typical drag formula (see section \ref{sec:astro}) may scale linearly with the gas parameter $\rho/c_{\rm s}^2$, implying that $\delta \phi_0\propto \rho /c_{\rm s}^2$. However, even within the same physical system, the variability can instead be determined by other combinations of parameters that better reflect the degree of non-linearity, turbulence, shocks, and etc.
\end{itemize}
Both of these aspects highlight the importance of modeling gas induced EE with sophistication, in order to avoid recovering inconsistent or even misleading results. At the same time, they show how dephasing in eccentric binaries could be used as a probe of the detailed non-linearity of gas flow around relativistic binaries, by allowing to recover individual Fourier modes of forces.
\section{Summary and Conclusion}
\label{sec:conclusion}
This work investigates eccentric binary systems evolving under gravitational wave (GW) emission and perturbed by external forces with complex Fourier spectra. Specifically, we presented a systematic way to dissect the components of the perturbative forces that produce secular drifts in the orbital elements, and therefore induce a dephasing of the GW signal. The main findings of our study are summarised below:
\begin{enumerate}
    \item We derived the GW phase shift components resulting from Fourier modes of the perturbative forces. These components are shown to introduce complex dephasing behavior that deviates significantly from the typical orbit averaged power-law approximations used in GW modeling. Our prescriptions therefore capture a broader and more nuanced range of physical scenarios.

    \item We demonstrated that even the simplest models of EE give rise to additional dephasing components due to resonances. These showcase frequency scalings that are distinct from what is usually associated to the respective EE. In the basic case where all Fourier coefficients are constant, we expect the $n$-th dephasing component $\delta \phi_n$ to be steeper by a factor $(f^{-19/18})^n$ with respect to the baseline $\delta \phi_0$. This has important consequences in the choice of dephasing prescriptions used in template matching for real eccentric signals, which are currently typically extrapolated from circular prescriptions.

    \item We provided strong evidence for the fact that the first epicyclical resonance will most often dominate the dephasing in gas embedded binaries with eccentricities larger than $\sim 0.05$, even in the equal mass case. Equal mass gas embedded binaries with a reference eccentricity of $e\sim 0.1$ will typically showcase larger dephasing by an \textit{order of magnitude} with respect to circular ones in the same environment. We show this by analysing a curated set of semi-analytical and numerical simulations of gas embedded binaries, building up from  linear dynamical friction to fully non-linear hydrodynamics.
\end{enumerate}
Our results highlight the high potential of eccentric binary sources of GWs in terms of scientific yield, but also the need for a more sophisticated treatment of EE.
Here we have, at a minimum, shown that eccentric sources embedded in gas will carry dephasing signatures of larger amplitude and steeper frequency scaling than what is typically considered. Furthermore, the amplitude of the dephasing will depend on gas properties that describe the variability of the gas flow, rather than its smoothed properties. These results have consequences in the study of EE for both equal mass ratio binaries (see e.g. \cite{2024garg,2025zwick}) and extreme mass ratio inspirals (see e.g. \cite{2022speri,2024duque,2025dyson}), which typically rely on a treatment of dephasing based on circular orbits.

Allowing instead to take a broader view, our results suggest that complex dephasing prescriptions, such as the ones discussed here, can be used to analyse the perturbative forces acting on inspiralling binaries with a surprising level of precision. Decomposing dephasing into its components allows to use eccentric GW sources as an unprecedented class of instrument, analogous to seismographs, able to probe the details of its surrounding environment by dissecting the various time-varying components of the coupling between binaries and their surroundings. With tentative evidence of eccentric signals already present in GW catalogues and  many works highlighting the importance of eccentricity for future detectors, such prospects showcase once more the unique potential of GW as a messenger.
\section*{Data Availability}
All simulation data, calculations and notebooks used to produce the results in this work will be shared upon reasonable request.
\begin{acknowledgments}
L.Z. acknowledges support from  ERC Starting Grant No. 121817–BlackHoleMergs. J.T. acknowledges support from the Horizon Europe research and innovation programs under the Marie Sk\l{}odowska-Curie grant agreement no. 101203883. 
C.T. acknowledges support by the European Union’s Horizon research and innovation program under Marie Sk\l{}odowska-Curie grant agreement No. 101148364.
D.ON, and D.J.D. acknowledge support from the Danish Independent Research Fund through Sapere Aude Starting Grant No. 121587.
The Center of Gravity is a Center of Excellence funded by the Danish National Research Foundation under grant No. 184.

\end{acknowledgments}


\appendix

\section{Higher order evolution equations}
We report here the evolution equations for the perturbed semi-major axis $a_{\rm p}$ and eccentricity $e_{\rm p}$ to order $\mathcal{O}(e^4)$, for use in applications with high eccentricity binaries. The general form is:
\begin{align}
\label{eq:adot_high}
    \dot{a}_{\rm p} \sqrt{\frac{GM}{p_0^3}} &= \sum_{n=0}^{4} A_n e_0^n \\
    \dot{e}_{\rm p}\sqrt{\frac{GM}{p_0}} &= \sum_{n=0}^{4} E_n e_0^n,
\end{align}
where the coefficients $A_n$ are given by:
\begin{align}
    A_0 &= 2 B_{0}^{\rm S} \\
    A_1 &= (i B^{\rm R}_1 + B_{1}^{\rm S})\\
    A_2 &= (i B^{\rm R}_2 + 2 B_{0}^{\rm S} + B_{2}^{\rm S})\\
    A_3 &=  (9 i B^{\rm R}_1 + 9 i B^{\rm R}_3 + 7 B_{1}^{\rm S} + 9 B_{3}^{\rm S})/8\\
    A_4 &=(5 i B^{\rm R}_2 + 4 (2 i B^{\rm R}_4 + 3 B_{0}^{\rm S} + B_{2}^{\rm S} + 2 B_{4}^{\rm S}))/6.
    \end{align}
Similarly we report the coeffcients $E_n$:
\begin{align}
    E_0 &=  (i B^{\rm R}_1 + 2 B_{1}^{\rm S})/2 \\
    E_1 &= (2 i B^{\rm R}_2 - 6 B_{0}^{\rm S} + 3 B_{2}^{\rm S})/4\\
    E_2 &= (-7 i B^{\rm R}_1 + 9 i B^{\rm R}_3 - 12 B_{1}^{\rm S} + 12 B_{3}^{\rm S})/16 \\
    E_3 &= (-7 i B^{\rm R}_2 + 8 i B^{\rm R}_4 - 10 B_{2}^{\rm S} + 10 B_{4}^{\rm S})/12 \\
    E_4 &= (34 i B^{\rm R}_1 - 621 i B^{\rm R}_3 + 625 i B^{\rm R}_5 + 60 B_{1}^{\rm S} \nonumber \\ &- 810 B_{3}^{\rm S} + 750 B_{5}^{\rm S})/768.
\end{align}
Note how higher order evolution in terms of eccentricity contains resonances from many force Fourier coefficients. The \texttt{Mathematica} notebook used to compute these coefficients to arbitrary order will be shared upon reasonable request, to spare the reader from immense tedium.
\bibliography{resonances}


\end{document}